\documentclass[twocolumn,preprintnumbers,amsmath,amssymb]{revtex4}
\usepackage{subfigure}
\usepackage{epsfig}

\usepackage{url}
\usepackage{graphicx}
\usepackage{dcolumn}
\usepackage{bm}
\usepackage{verbatim}
\usepackage{subfigure}
\usepackage{rotating}
\usepackage{afterpage}

\newcommand{\pt}{$p_{T}$ }
\newcommand{\et}{$E_{T}$ }

\begin{document}
\title{Jet - Underlying Event Separation Method for Heavy Ion Collisions at the Relativistic Heavy Ion Collider}
\author{J.~A.~Hanks$^1$, A.~M.~Sickles$^2$, B.~A.~Cole$^3$, A.~Franz$^2$, M.~P.~McCumber$^4$, 
D.~P.~Morrison$^2$, J.~L.~Nagle$^4$, C.~H.~Pinkenburg$^2$, B.~ Sahlmueller$^1$, P.~Steinberg$^2$, 
M.~von Steinkirch$^1$, M.~Stone$^4$}
\affiliation{$^1$ Department of Physics and Astronomy, Stony Brook University, SUNY, Stony Brook, New York 11794-3400, USA}
\affiliation{$^2$ Physics Department, Brookhaven National Laboratory, Upton, New York, 11973-5000}
\affiliation{$^3$ Columbia University, New York, New York 10027 and Nevis Laboratories, Irvington, New York 10533, USA}
\affiliation{$^4$ University of Colorado, Boulder, Colorado 80309, USA}
\date{\today}
\begin{abstract}
Reconstructed jets in heavy ion collisions are a crucial tool for understanding the quark-gluon plasma.  The separation
of jets from the underlying event is necessary particularly in central heavy ion reactions in order to quantify medium
modifications of the parton shower and the response of the surrounding medium itself.  
There have been many methods proposed and implemented
for studying the underlying event substructure in proton-proton and heavy ion collisions.  In this paper, we detail
a method for understanding underlying event contributions in Au+Au collisions at $\sqrt{s_{NN}}$ = 200  GeV utilizing the
HIJING event generator~\cite{Gyulassy:1994ew}.   This method, extended from previous work by the ATLAS 
collaboration~\cite{atlas_note},
provides a well-defined association of ``truth jets'' from the fragmentation of hard partons with ``reconstructed jets'' 
using the anti-$k_T$ algorithm. 
Results presented here are based on an analysis of 750M minimum bias HIJING events.
 We find that there is a substantial range of 
jet energies and radius parameters where jets are well separated from the background fluctuations (often termed ``fake jets'') 
that make jet measurements at RHIC a compelling physics program.
\end{abstract}
\maketitle
\section{Introduction}
Understanding the detailed interaction and coupling between hard scattered partons and the quark-gluon plasma through which
they propagate is essential to our fundamental knowledge of QCD and in determining properties of the quark-gluon plasma.  The 
measurement of fully reconstructed jets in heavy ion collisions at the LHC~\cite{Aad:2010bu,Chatrchyan:2011sx} 
highlight the substantial
additional information contained therein and its complementary nature to single 
hadron~\cite{Adcox:2001jp,Adams:2003kv,Adare:2008qa}, 
di-hadron correlations~\cite{Adams:2006yt,Adare:2008cqb,Abelev:2009qa,Adare:2010ry}.  
The measurement of direct photon-jet correlations is another critical handle 
to be utilized~\cite{Wang:1996yh}.
Extending fully calorimetric jet 
measurements to lower center-of-mass energies at the Relativistic 
Heavy Ion Collider 
provides measurements for kinematics difficult to access at the LHC and the QGP at different temperature and
coupling regime. 

With the first Pb+Pb at $\sqrt{s_{NN}}$=2.76 TeV collisions at the LHC new insights into jet physics in heavy ion collisions 
were gained.  The ATLAS collaboration reported an increase in the number of energy asymmetric di-jets
in central Pb+Pb collisions compared to proton-proton and peripheral Pb+Pb collisions~\cite{Aad:2010bu}.  They also 
reported
the suppression of jets with 100$<p_T<$200  GeV/c by a factor of 
approximately when comparing central to peripheral Pb+Pb collisions~\cite{Steinberg:2011qq}.  
The CMS collaboration measured jet-hadron correlations in a similar jet \pt
range and found that the energy lost by high \pt fragments was approximately balanced by
very low \pt tracks far from the jet axis~\cite{Chatrchyan:2011sx}.
However the data from both RHIC and the initial LHC results are not enough to constrain the 
physics of jet quenching.  Most theoretical descriptions have relied on weakly coupled
techniques~\cite{Majumder:2010qh}.  Features of strong coupling, as observed in descriptions
of the bulk matter, might contribute to jet quenching as well.  More data on jet 
observables (including dijet, $\gamma$-jet and heavy flavor tagged jets) at RHIC and the
LHC will be necessary to understand the physics of jet quenching over the full range of medium properties and
jet kinematics and probe for sensitivity of the quenching to outgoing parton virtuality.

The multiplicity of charged particles $dN_{ch}/d\eta$ is approximately 2.15 times higher for Pb+Pb central collisions 
at $\sqrt{s_{NN}}$ = 2.76 TeV compared with Au+Au central collisions at $\sqrt{s_{NN}}$=200  GeV~\cite{Aamodt:2010cz}.  Thus the
soft particle background is substantially higher for LHC events.  However, the jet cross section is substantially higher as well, 
and measurements for jets with energies greater than 100  GeV appear well separated from the background (though 
detailed publications of these studies are not yet available).  Various methods have been explored at the LHC and RHIC
for understanding the underlying event contributions, and what are often referred to as 
``fake jets''~\cite{Lai:2009ai,Jacobs:2010wq,atlas_note,Cacciari:2011tm,Abelev:2012ej}.  

At $\sqrt{s_{NN}}$=200 GeV the projected jet rates into $|\eta|<$1 based on NLO pQCD cross sections~\cite{Vogelsang:NLO} and
expected RHIC luminosities have been computed~\cite{decadal_plan}.  In a typical year of RHIC running 50B Au+Au events
could be sampled.  In the top 20\% centrality that would lead to approximately $10^7$ jets above 20 GeV, $10^6$ jets
above 30 GeV, $10^5$ jets above 40 GeV and $10^4$ jets above 50 GeV.  Over 60\% of the time there is full containment of the 
opposing dijet for 20  GeV jets with that percentage increasing with increasing jet energy. 

In this paper, we present a study of jet reconstruction and separation from the underlying event using the  
HIJING~\cite{Gyulassy:1994ew} event generator for Au+Au events at $\sqrt{s_{NN}}$=200  GeV.  This follows an iterative 
underlying event subtraction procedure extended from one developed by the ATLAS Collaboration~\cite{atlas_note}.  While the
exact definition of a correctly reconstructed jet versus a ``fake jet'' is arbitrary, this methodology allows us to make a
well-defined and documented comparison to cross-check with other methods.

\begin{figure*}
\centering
\includegraphics[width=0.8\textwidth]{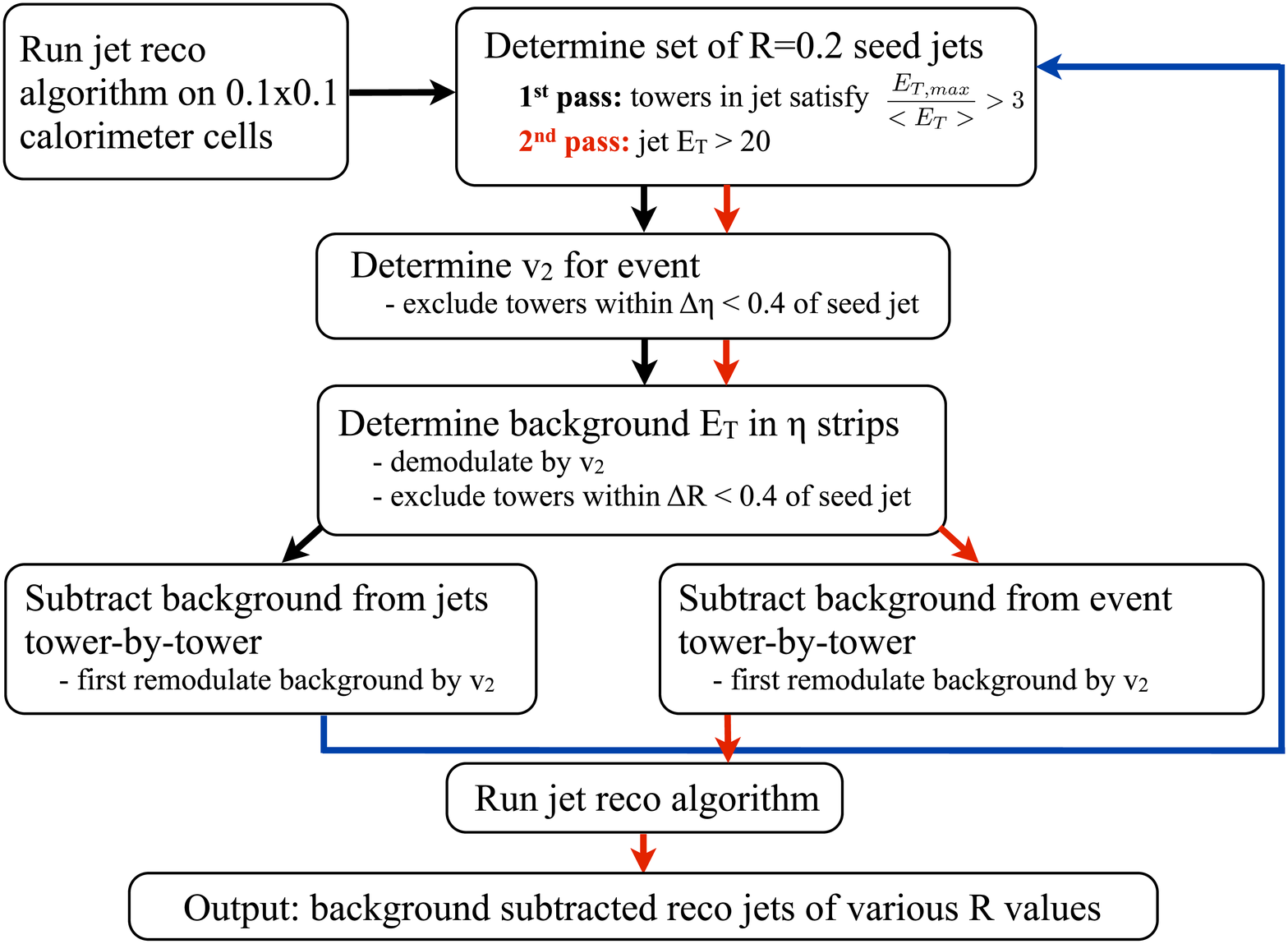}
\caption{Schematic illustration of the jet background subtraction method.}
\label{sub_scheme}
\end{figure*}

\section{Jet - Underlying Event Separation Methodology}
\label{sec:method}

For these studies we utilize the HIJING (version 1.383) event generator run with standard settings and
quenching turned off for Au+Au collisions
at $\sqrt{s_{NN}}$=200  GeV.  
HIJING is a QCD based Monte Carlo for the study of jet production in high energy nucleus-nucleus collisions.
For these initial studies, we 
explore what a ``perfect'' detector is capable of measuring.  We assume a segmentation in $\Delta \eta \times \Delta \phi$ = $0.1 \times 0.1$ and that all particle energies are recorded perfectly (with the exception of neutrinos and muons).  We assume
a nominal coverage of $|\eta| < 1.0$ and full azimuthal coverage.  For the entire study we utilize the anti-$k_T$ jet
reconstruction algorithm~\cite{Cacciari:2008gp} (part of the
FastJet package~\cite{Cacciari:2005hq}) with radius parameters R=0.2, 0.3, and 0.4.

A schematic diagram of the underlying event subtraction steps is shown in Fig.~\ref{sub_scheme}.
The first step is to run the anti-$k_{T}$ algorithm over the full set of energy values (unsubtracted) with R=0.2 
and record the jet axis coordinates in $\eta$ and $\phi$.  
This initial suite of jet candidates is used to exclude regions around these jets from the 
initial underlying event average energy. 
Exclusion regions are defined by R=0.2 jets  in which the maximum tower in the jet has an energy of more than 
three times the average tower energy in the jet.
We then exclude all energies for $0.1 \times 0.1$ cells whose center coordinate is within 
$\Delta R=\sqrt{\Delta\eta^{2}+\Delta\phi^{2}} < 0.4$ 
of any of the above initial jet candidates.  The remaining energy values are 
used to determine the average cell energy (i.e. in the non-jet regions) in $\Delta\eta=$ 0.1 strips. The modulation in the background 
due to flow must first be removed, so the $\left< cos(2\phi) \right>$ (i.e. $v_{2}$ parameter) is also determined for the energy 
distribution and removed from each cell before determining the average. Only the $\eta$ strips which have complete 
$\phi$ coverage after the determination of the exclusion regions are used in the $v_2$ determination.

The HIJING generator has no bulk collective flow and thus has only a modest $\left< cos(2\phi) \right>$ from 
decays, di-jet correlations and global momentum conservation.  
As the flow modulation of the underlying event is an important component of any
subtraction procedure on real data, we have added a flow modulation to the individual HIJING particles prior to 
segmenting the energies into cells.  The flow parameterization~\cite{Masera:2009zz}
 is based on fits to the available data.
Higher flow moments have an increasing relative importance for more central events~\cite{Adare:2011tg}, and can be incorporated
in future studies.

This underlying event average energy is a zeroth order estimate since
the initial jet determination does not have an underlying event subtraction.
We now subtract the $v_2$ modulated underlying event energy cell-by-cell from the cells contained by the initial 
set of R=0.2 jets to get a better estimate of the jet $E_T$.
At this second iterative step, new exclusion regions are defined by towers with
$\Delta R<$ 0.4 around background subtracted jets with $E_T>$ 20  GeV.  The underlying event and $v_2$ are 
re-determined removing towers within $\Delta \eta<$0.4 of the jets
 as described above and the background re-subtracted from the original unsubtracted towers.  
Finally the anti-$k_T$ jet algorithm is run on the background subtracted towers with a range of R values (0.2, 0.3, 0.4).

When the jet reconstruction is run over background subtracted towers many of the towers
have negative $E_T$.  We modify these towers to have a small positive energy before passing them
to the jet reconstruction algorithm.  After the towers are grouped into jets we recalculate the
jet $E_T$ including the negative energy.

\section{HIJING Truth Information}

In order to identify ``true jets'' from the HIJING event generator, we have augmented the code so that every
time the fragmentation routine (HIJFRG) is called, we record the set of final state hadrons that result from that fragmentation.  
We then run the anti-$k_T$ algorithm on those final state hadrons (using their exact momentum vectors).
The jet reconstruction is run once for each anti-$k_{T}$ jet R parameter under consideration and the resulting
``true jet'' information is added to the output.

Before presenting the results, it is important to define our terms.  Even in a model such a HIJING where all truth
information is known, there is an arbitrariness in the definitions of ``fake jets'' and ``true jets'' as examples.  
For example, consider a HIJING fragmentation call that results in hadrons reconstructed via anti-$k_T$ with R = 0.4 
into a jet with an energy of 20  GeV. 
If after running jet reconstruction over the full HIJING event
 one reconstructs a jet using anti-$k_T$ with R = 0.4 that has a jet axis
within $\Delta R \equiv\sqrt{\Delta\eta^{2}+\Delta\phi^{2}} < 0.25$ and energy 18  GeV, common sense might dictate that this was a ``true jet''
and the 2  GeV difference is a result of the fluctuations in the underlying event.  However, imagine that the HIJING
fragmentation reconstructed to an energy of 4  GeV and the full HIJING event resulted in a jet along the same axis
but with energy of 40  GeV.  Common sense might dictate that this was a ``fake jet'' (i.e. a very small jet that combined with 
substantial background fluctuations that results in a very large energy reconstruction).  

Here, we define a ``fake jet'' as one where the associated HIJING fragmentation jet is less than 5  GeV 
(or does not exist at all).  The jet is a good ``true jet'' if there is an associated HIJING fragmentation jet 
within $\Delta R =\sqrt{\Delta\eta^{2}+\Delta\phi^{2}} < 0.25$ and greater than 5  GeV.  We then examine in detail the HIJING
fragmentation jet energy distribution for those associated with different selected fully reconstructed jet energies.  
In principle, one could introduce no such arbitrary definition and put everything into a response matrix down to the lowest
energy scales.  In practice, if there are substantial contributions of very low energy HIJING fragmentation jet
energies to high energy reconstruction jets it will be nearly impossible to control the systematics and unfold such a
matrix.

Results presented here are based on an analysis of 750M minimum bias HIJING events.

\section{Results}

\begin{figure}
\centering
\includegraphics[width=\columnwidth]{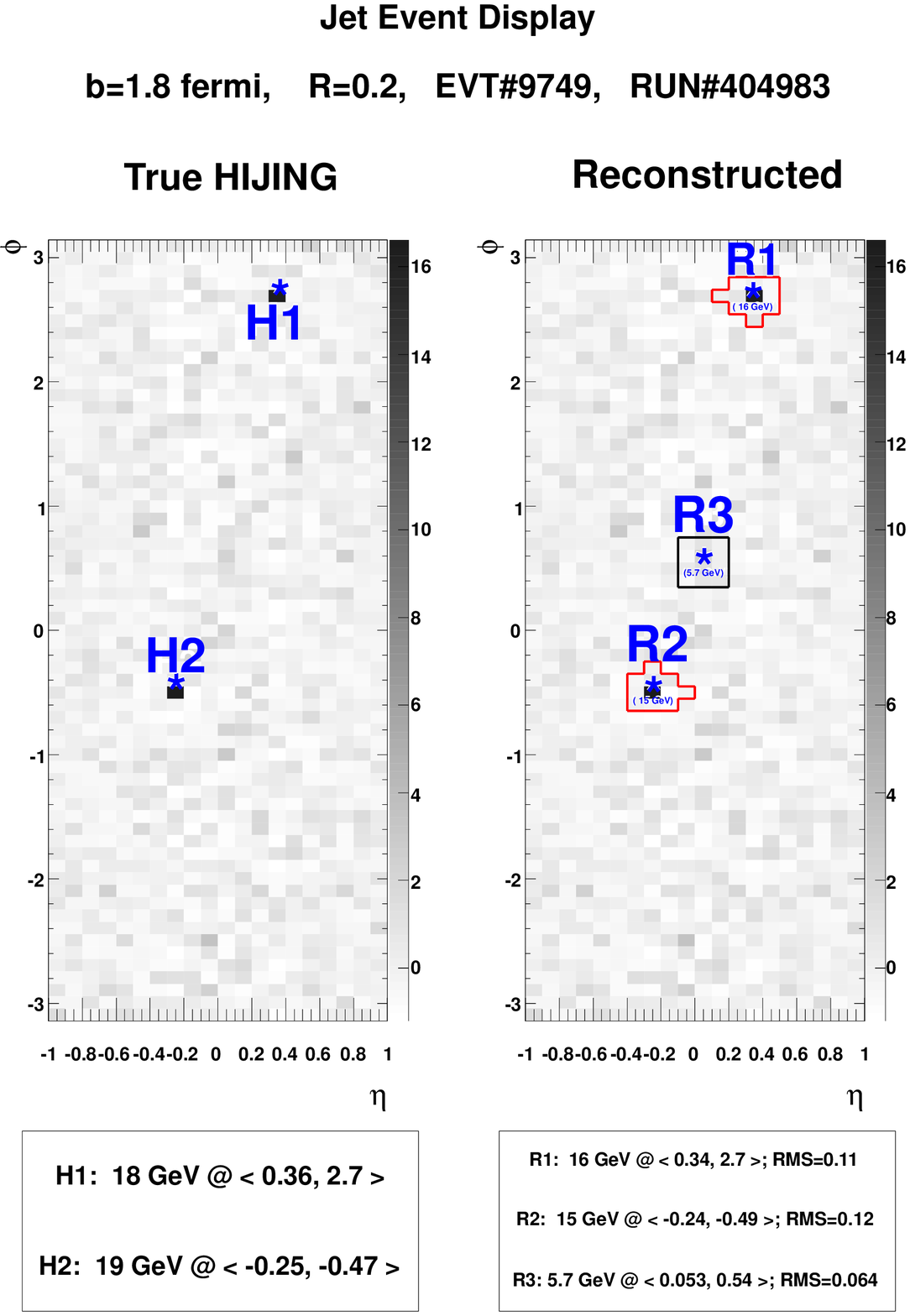}
\caption{Event display for a $E_T=$ 18  GeV true dijet pair jet matched to a $E_T$=15 and 16 GeV reconstructed jets
in a b=1.8fm HIJING event.  
All jets shown
in this event display are reconstructed with the anti-$k_T$ algorithm with R=0.2.  Both histograms show the
background subtracted 0.1x0.1 $\eta-\phi$ tower energy.  A minimum $E_T$ cut of 5  GeV is placed on all jets
shown in this display. The stars in the left panel show the true HIJING jets  and box below shows
jet \et and $\eta$,$\phi$ location of the jet axis.  The right panel shows the reconstructed jets.  The jet
labeled R1 is reconstructed at $E_T=$16  GeV and matched to the H1 jet in the left panel and R2 is
matched to H2.  The other
reconstructed jet in the event with $E_T>$ 5  GeV is shown as R3.  It is not associated with any true jets
with $E_T>$ 5  GeV.}
\label{real_jet_r2}
\end{figure}

\begin{figure}
\centering
\includegraphics[width=\columnwidth]{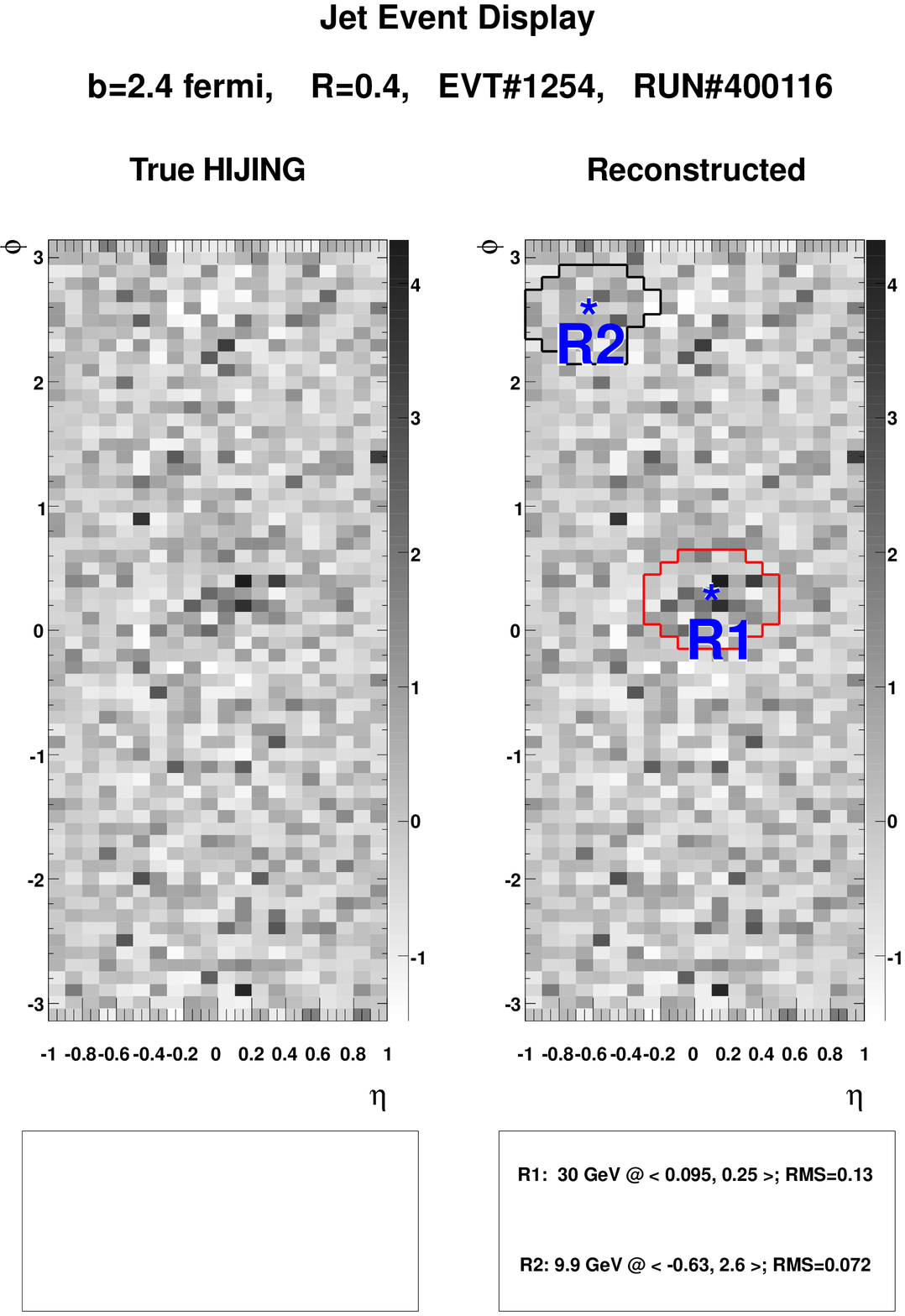}
\caption{Event display for a $E_T=$ 30  GeV fake jet.  All jets shown
in this event display are reconstructed with the anti-$k_T$ algorithm with R=0.4.  Both histograms show the
background subtracted $0.1 \times 0.1$ $\eta-\phi$ tower energy.  A minimum $E_T$ cut of 5  GeV is placed on all jets
shown in this display.  There are no true HIJING jets with $E_T>$5 GeV in this event. 
  The right panel shows the reconstructed jets.  The jet
labeled R1 is reconstructed at $E_T=$ 30  GeV.  The other
reconstructed jet in the event with $E_T>$ 5  GeV is shown as R2.  
 The jet \et and $\eta$,$\phi$ locations are shown in the bottom right box.}
\label{fake_jet_r4}
\end{figure}

In order to illustrate the background subtraction procedure, we show a selection of event displays. 
Figure~\ref{real_jet_r2} shows a true dijet pair with R = 0.2 where both jets have been matched
to reconstructed jets.  The
reconstructed jet has an axis within $ \Delta R < $ 0.1 of the true. 
 Also shown in the event are the next highest \et reconstructed jet.  This
jet is not matched to any true jets with $E_T>$ 5  GeV and has $E_T$ in the region 
where we expect fake jets to dominate.  Figure~\ref{fake_jet_r4} shows a fake jet with $E_T = $30  GeV which is
not matched to any true jet from the HIJING event.  One other fake jet, also not matched to any true jets, 
is shown on the
plot.  

\begin{figure}
\centering
\includegraphics[width=\columnwidth]{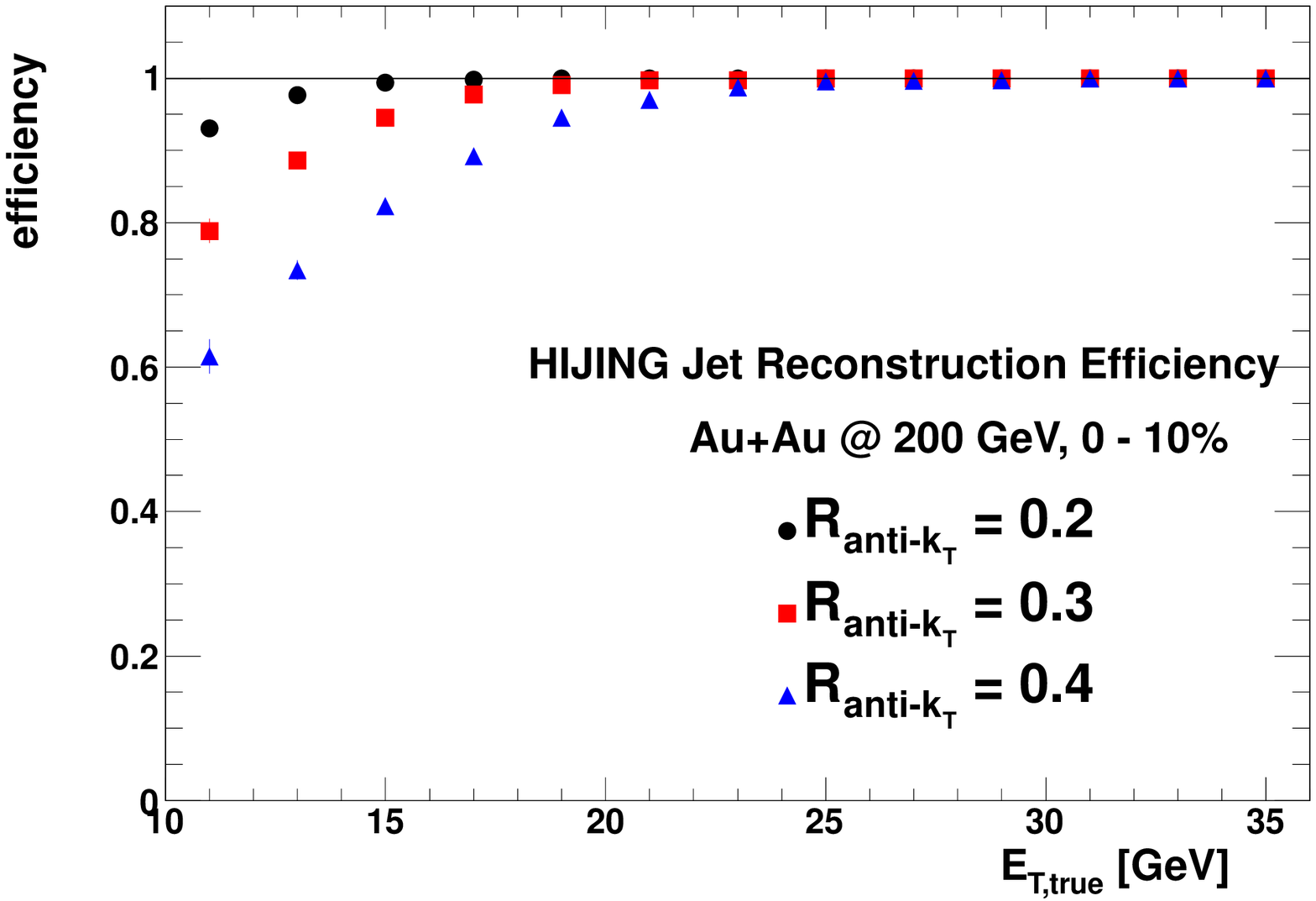}
\caption{Matching efficiency for true HIJING jets as a function of jet \et for anti-$k_T$ R parameters 0.2 (black),
0.3 (red) and 0.4 (blue). For a jet to be considered matched the reconstructed jet axis must
be within $\Delta R<$0.25 of the HIJING jet axis.}
\label{jet_match_eff}
\end{figure}

We concentrate on central collisions where the underlying event background is largest.
For this study we define collision centrality in the HIJING events by the number of charged particles
with psuedorapidity 3$<\eta<$4.
Figure~\ref{jet_match_eff} shows the efficiency of finding matches to true jets in the most central 10\% of collisions
for the various anti-$k_T$ R parameters as a function of the true jet $E_T$.  For all R parameters
the efficiency rises with jet $E_T$ and approaches 100\% between 20 and 30 GeV.

\begin{figure*}
\subfigure[]{
\includegraphics[width=0.30\textwidth]{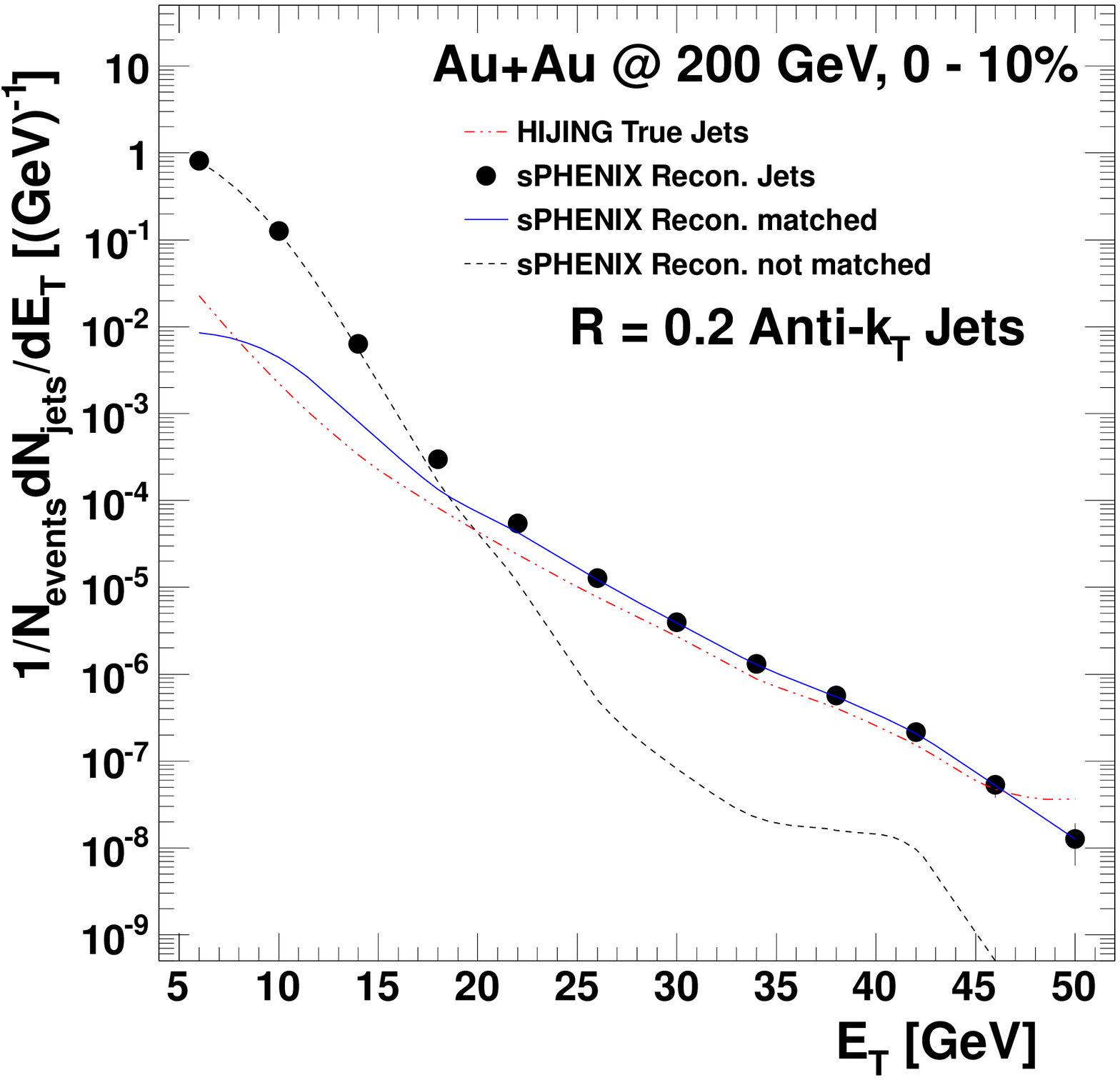}
\label{fake_fig2}}
\hspace{0.02\textwidth}
\subfigure[]{
\includegraphics[width=0.30\textwidth]{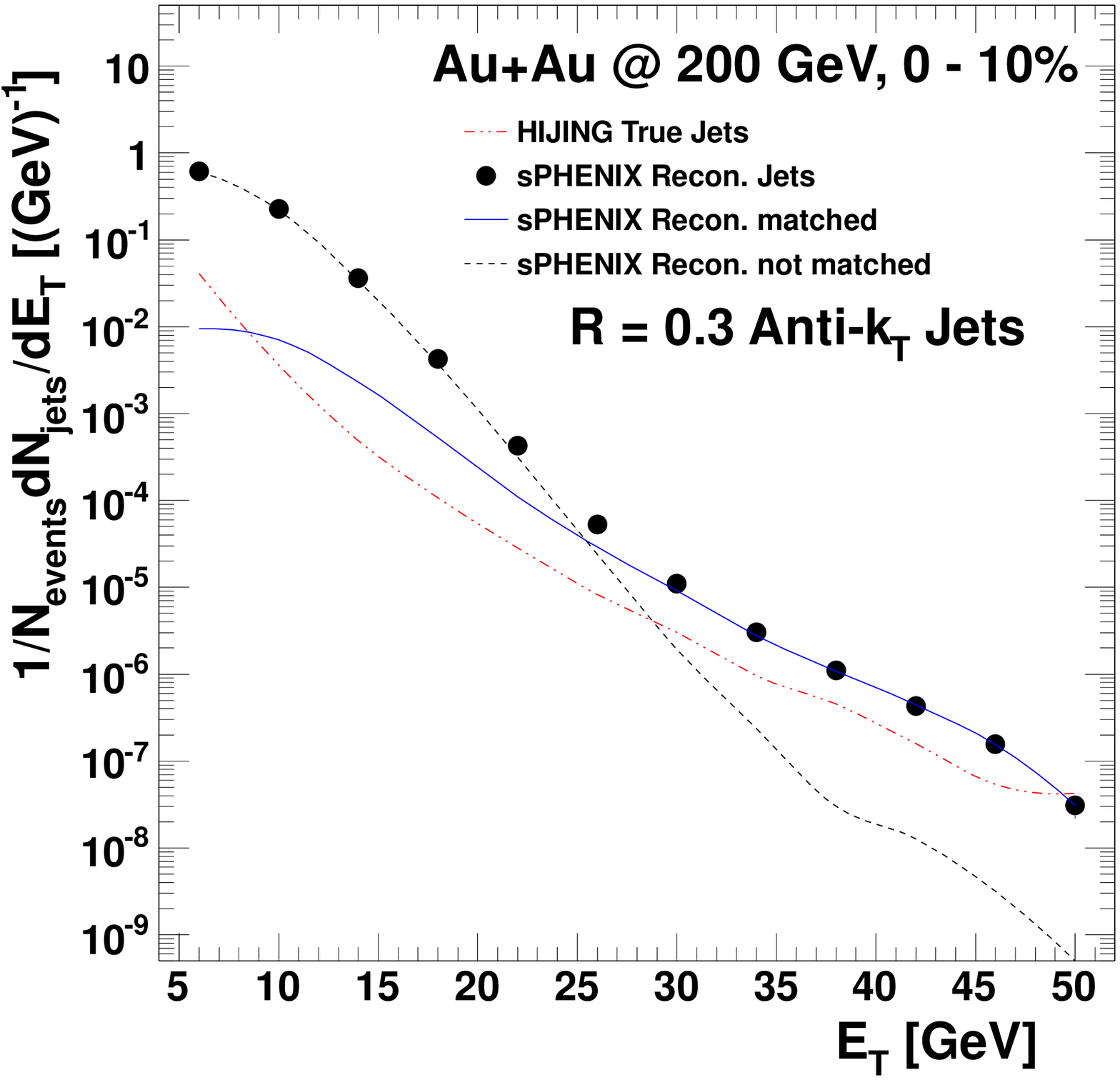}
\label{fake_fig3}}
\hspace{0.02\textwidth}
\subfigure[]{
\includegraphics[width=0.30\textwidth]{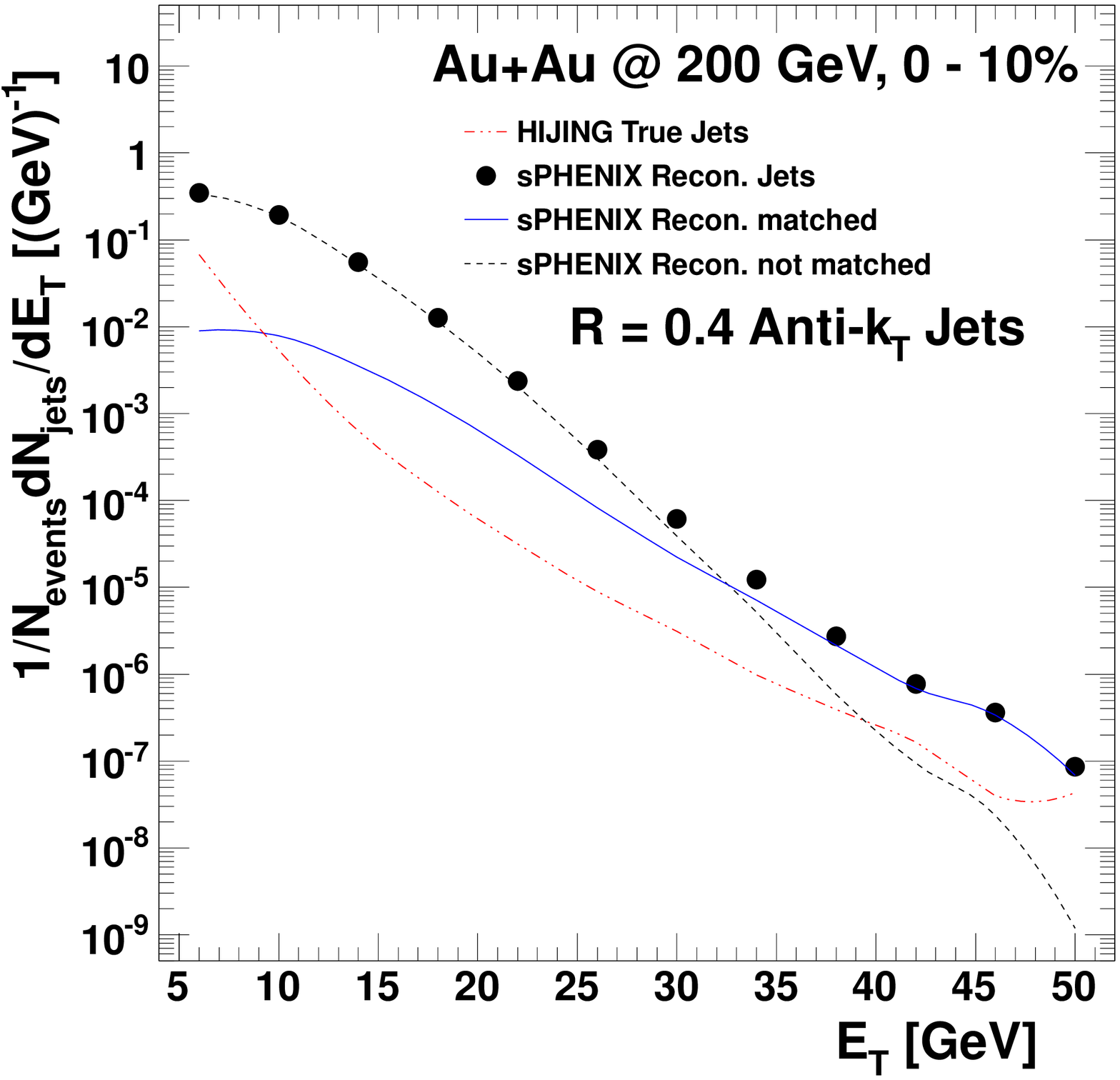}
\label{fake_fig4}}
\caption{\et spectra for true HIJING jets (red line) and reconstructed jets (black points).
The reconstructed jets are further divided into those which are matched to a true HIJING jet (blue line) and those
which are not matched to a true HIJING jet (``fake jets'', black line).  To be considered matched
the axis of the true HIJING jet and the reconstructed jet must be within $\Delta R <$0.25 and
the HIJING jet must have $E_T >$5 GeV. Shown are results for 0-10\% central HIJING events
using anti-$k_{T}$ jets with R=0.2 (a), R=0.3 (b) and R=0.4 (c).}
\label{fake_fig}
\end{figure*}

However, in order to quantify the jet performance we need to understand the contribution to the reconstructed jet
\et spectrum from jets which are not matched to any true HIJING jet, ``fake jets''.
In Figure~\ref{fake_fig} we show the true, reconstructed and fake jet
\et spectra for R = 0.2 (left), 0.3 (middle) and 0.4 (right) for the 10\% most central Au+Au at $\sqrt{s_{NN}}$= 200  GeV
HIJING events.  
Shown as red are the true HIJING fragmentation jet distributions.  
The points show the final reconstructed jet distribution.  This is broken down into those jets that are matched with a 
true HIJING jet  and those that are not matched with a true HIJING jet.  
To be considered matched the jet axis of the reconstructed jet must be within $\Delta R < $ 0.25 of the true HIJING jet
and the HIJING jet must have $E_{T}>$5 GeV.
One observes a good match between true HIJING and matched reconstructed jet distributions taking into account the
additional energy resolution blurring from the underlying event subtraction.  
One observes a very large contribution fraction of reconstructed jets are not matched at low \et; the fraction
then falls quickly and goes below the matched reconstructed jets at around 18  GeV in the R = 0.2 case.  The crossing
point is at higher \et for R=0.3 and 0.4 jets, around 25 and 30  GeV, respectively.

\begin{figure*}
\centering
\subfigure[]{
\includegraphics[width=0.45\textwidth]{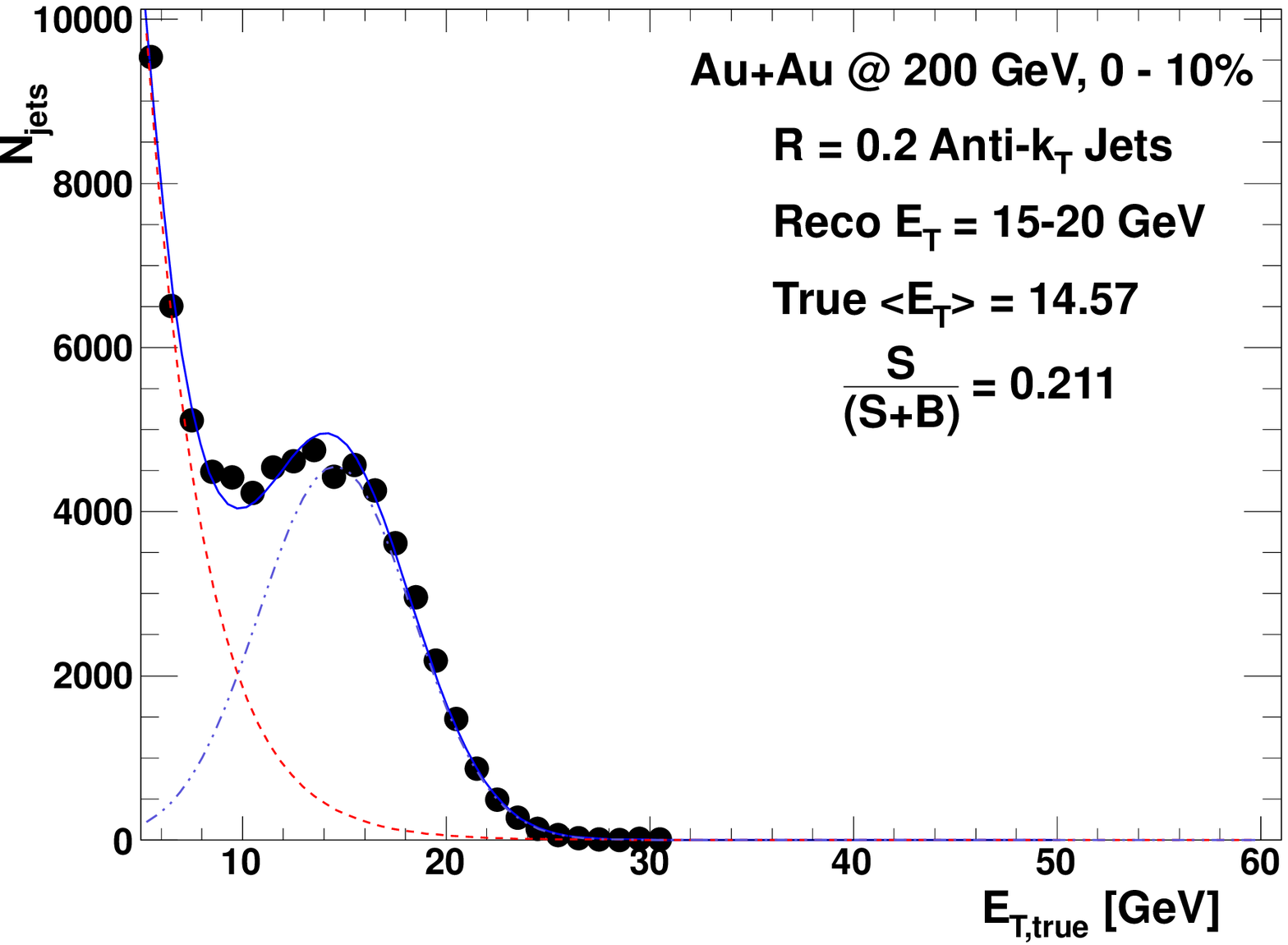}}
\hspace{0.05\textwidth}
\subfigure[]{
\includegraphics[width=0.45\textwidth]{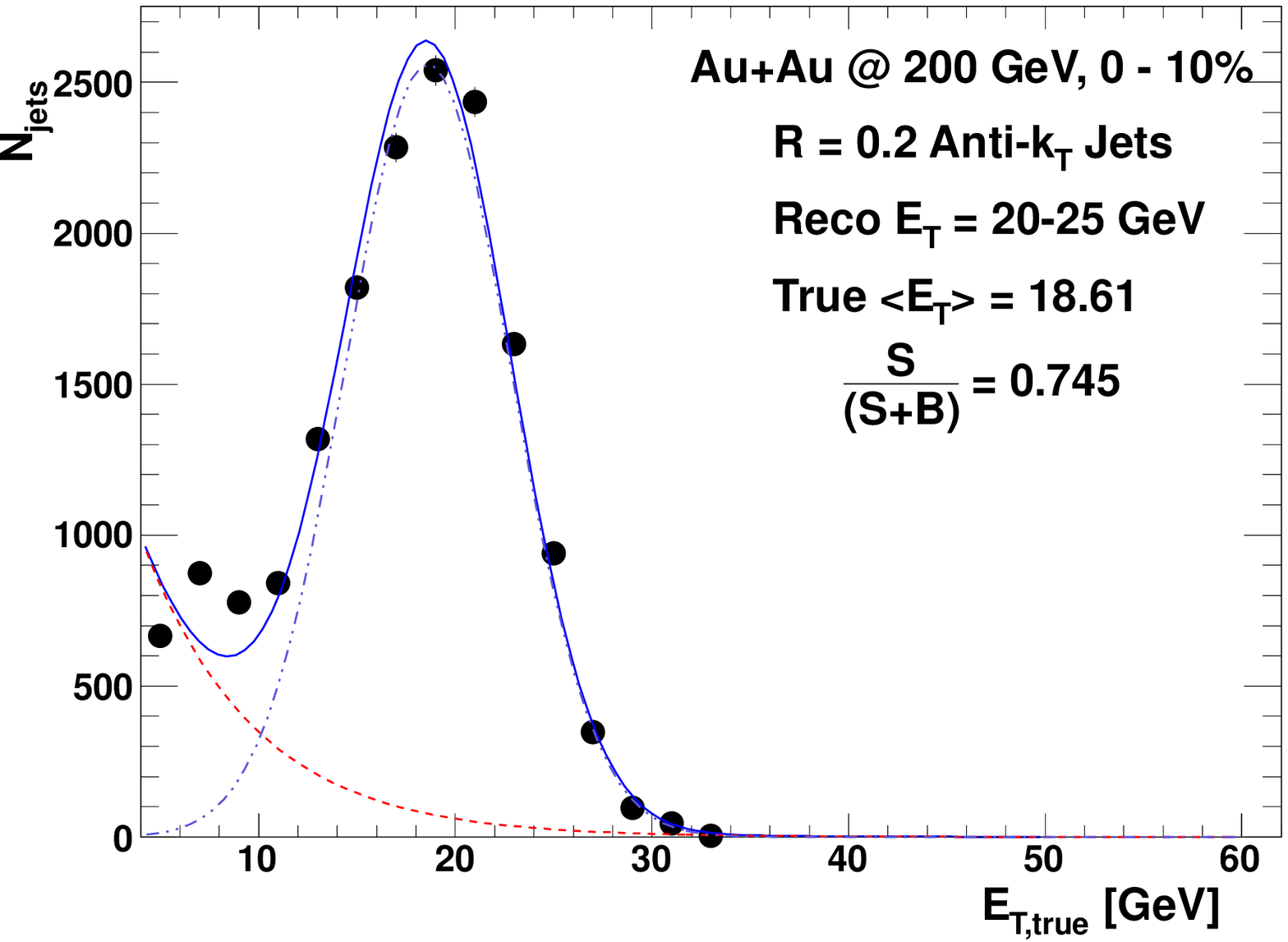}}
\subfigure[]{
\includegraphics[width=0.45\textwidth]{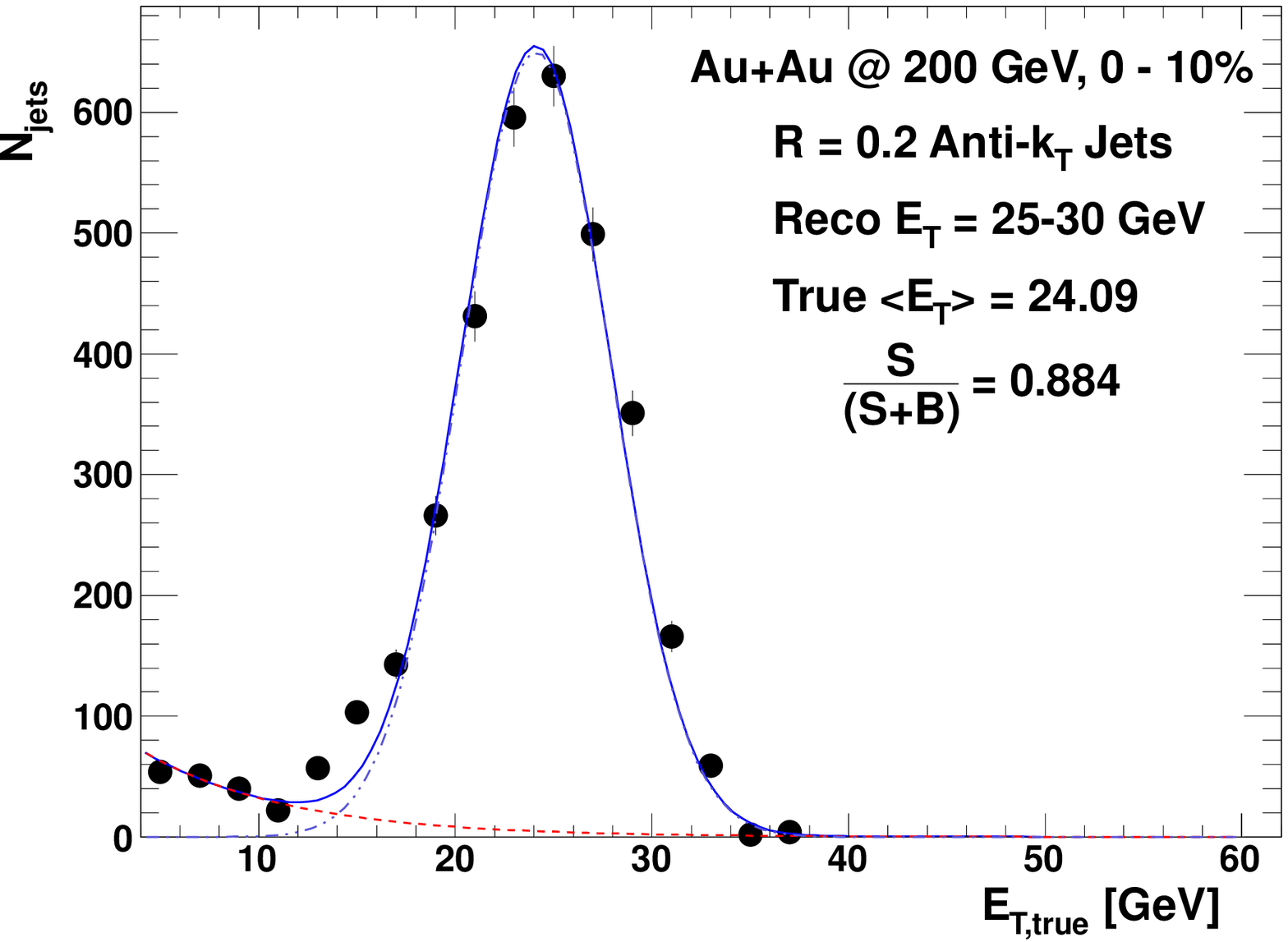}}
\hspace{0.05\textwidth}
\subfigure[]{
\includegraphics[width=0.45\textwidth]{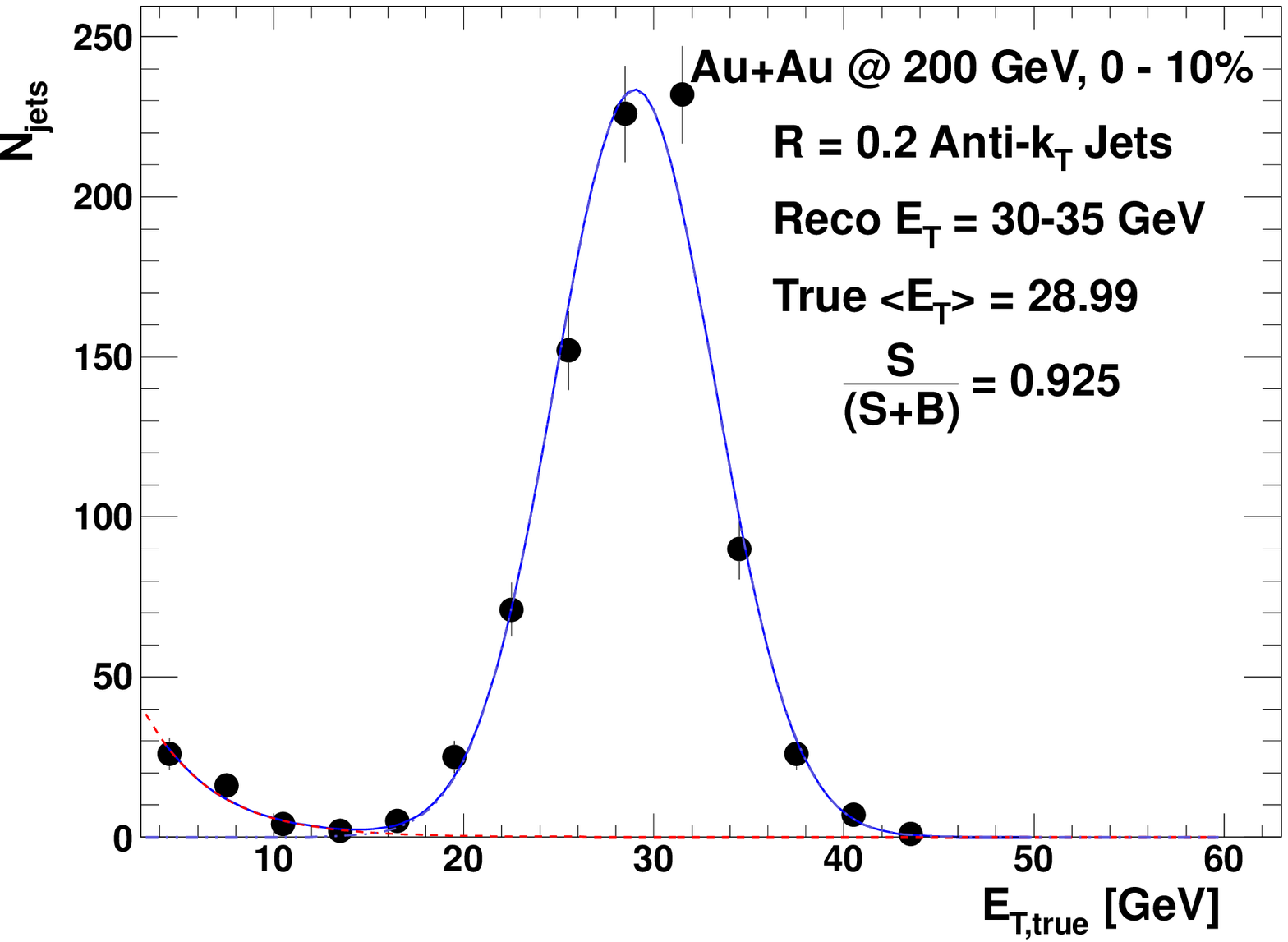}}
\caption{True \et for reconstructed jets anti-$k_{T}$ R = 0.2 for the reconstructed jet \et 
15-20 GeV (a), 20-25 GeV (b), 25-30 GeV (c) and 35-40 GeV (d).
The lines show the results of fits containing a background component which is exponentially 
falling (dashed line) and a signal Gaussian component (dot-dashed line).  The total fit is
shown as a solid line.  The plots show the $\frac{S}{S+B}$ where the signal (S) is determined from the area
under the Gaussian within $\pm$2$\sigma$ of the mean and the total background (B)
includes both those jets reconstructed with a $>2\sigma$ energy mismatched and those 
which were not matched at all to a HIJING jet. Fit parameters are shown in Table~\ref{tab2}.}
\label{resp2}
\end{figure*}

The low fake rate alone is not sufficient due to the arbitrariness of the 5  GeV separation between ``fake'' and ``true'' 
jet fragmentation associations.  Shown in Figure~\ref{resp2} is the distribution of HIJING true energies 
for fully reconstructed jet energies (with different selections).  The upper left panel for reconstructed jets with 
R = 0.2 and energies 15-20  GeV shows a peaked distribution around $\approx$ 15  GeV.   
The tail to lower energies could in principle be accounted for in a response matrix (though with great care and systematic cross checks).  
However, as one moves to higher energies 25-30  GeV, there is essentially no tail contributions and a peak around 26 GeV
 and width of 5  GeV.  This indicates
a regime where a standard response matrix and unfolding procedure should be successful.  
Similar plots are shown for R = 0.3 and R = 0.4.
There is a shift downward from the reconstructed jet energies to the corresponding true jet energies due
to the rapid fall off of the jet cross section with energy and a tail to low HIJING jet energies that disappears with
increasing reconstructed jet \et and the corresponding decrease of fake jets.

In order to quantify the purity of the reconstructed jet sample,
we have fit the distributions with a background contribution which falls
exponentially with increasing jet \et and a Gaussian with a free mean and width.  The results of those
fits, along with the fractions of the total reconstructed jets (both matched and unmatched)
 which are included in the Gaussian are shown in 
Tables~\ref{tab2}-\ref{tab4}.

\begin{figure*}
\centering
\subfigure[]{
\includegraphics[width=0.45\textwidth]{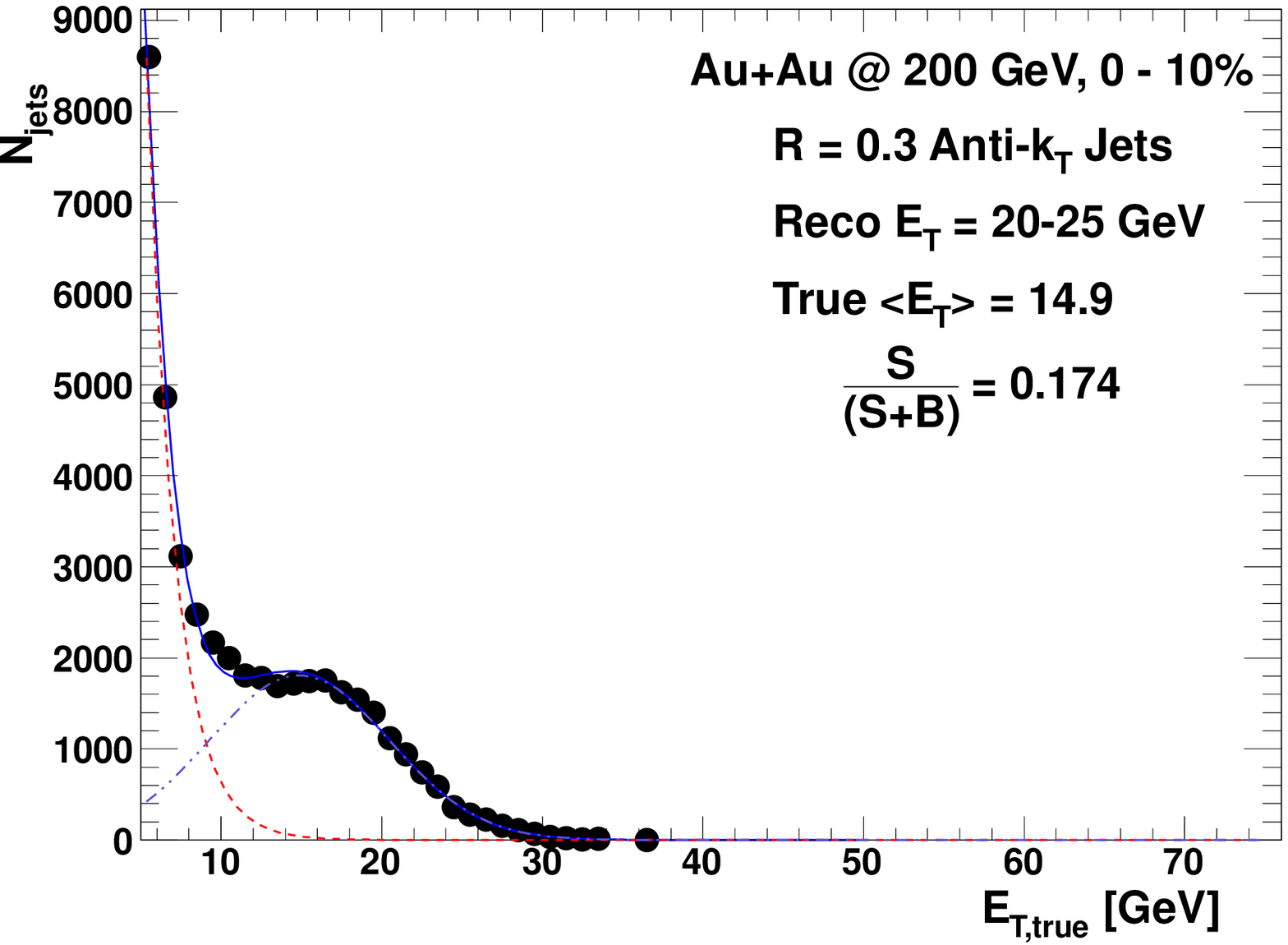}}
\hspace{0.05\textwidth}
\subfigure[]{
\includegraphics[width=0.45\textwidth]{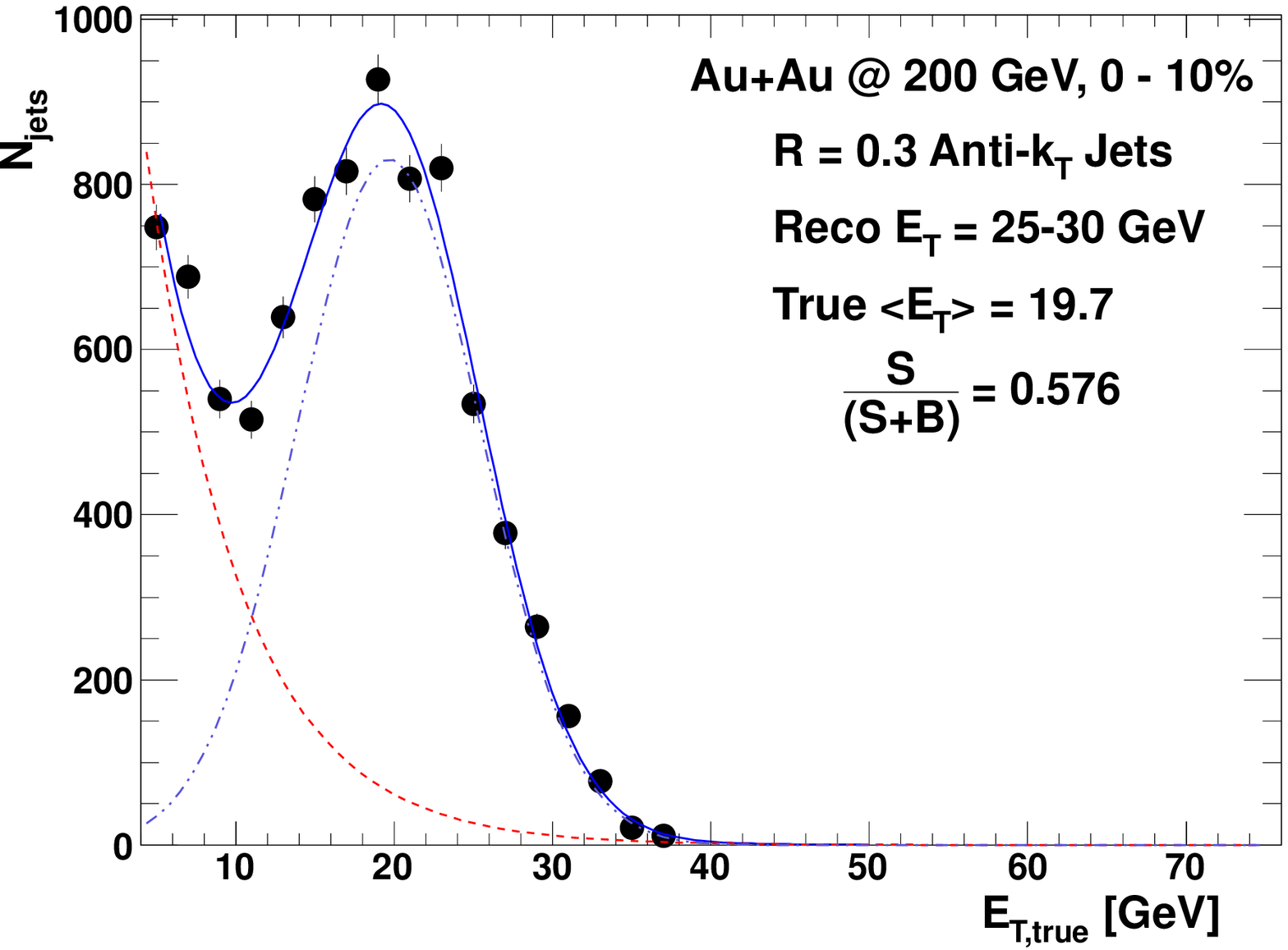}}
\subfigure[]{
\includegraphics[width=0.45\textwidth]{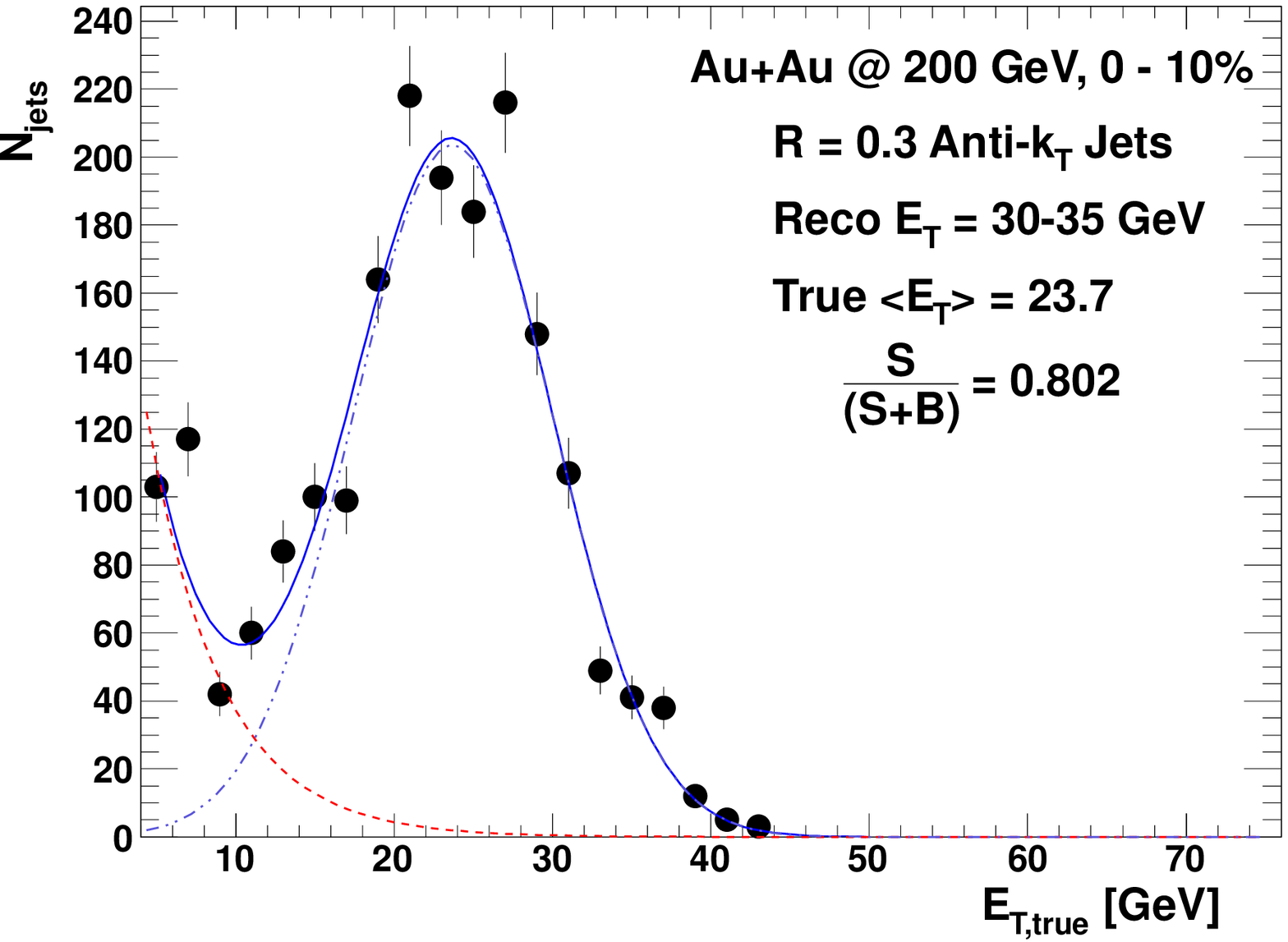}}
\hspace{0.05\textwidth}
\subfigure[]{
\includegraphics[width=0.45\textwidth]{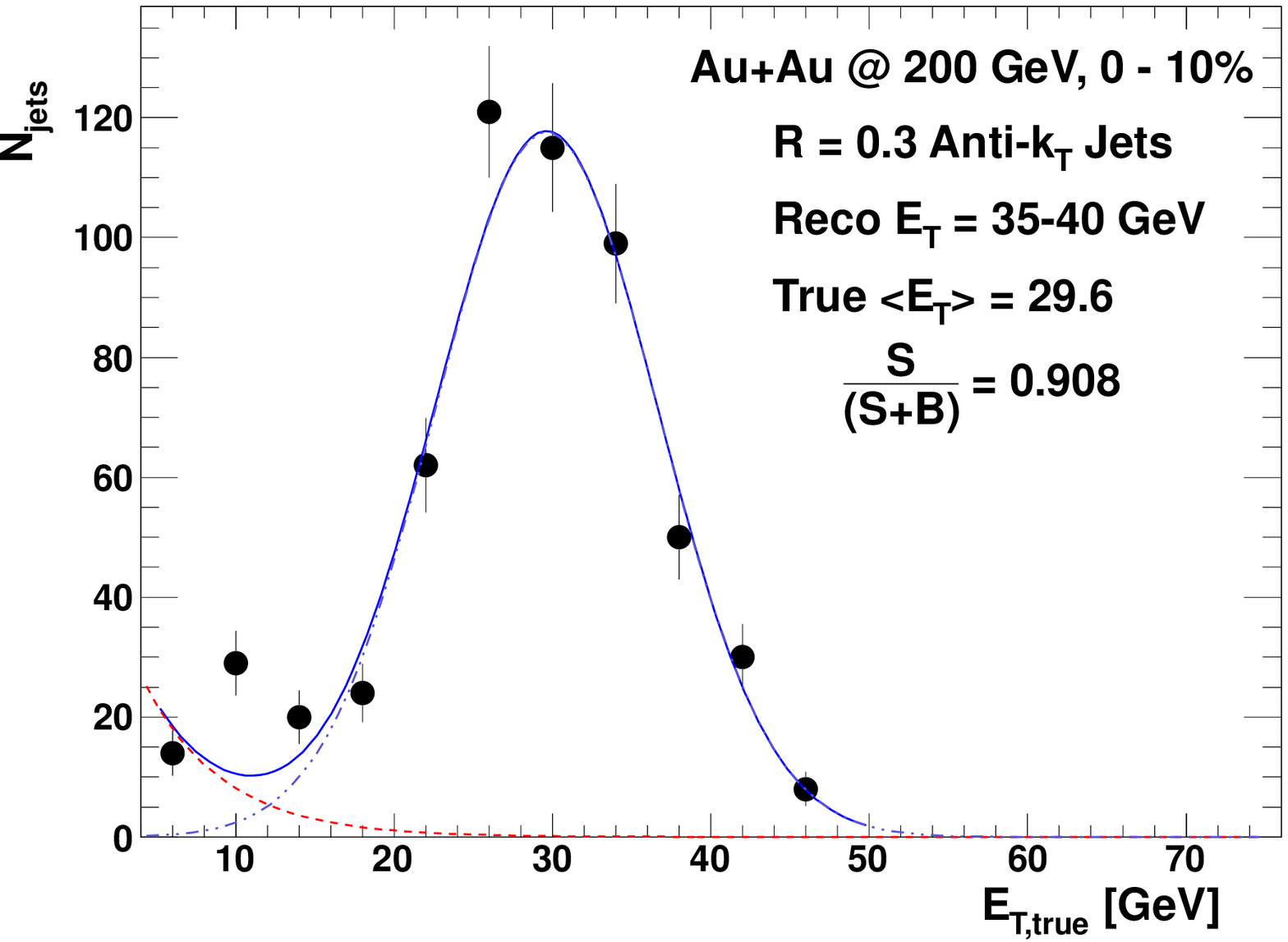}}
\caption{True \et for reconstructed jets anti-$k_{T}$ R = 0.3 for the reconstructed jet \et 
20-25 GeV (a), 25-30 GeV (b), 30-35 GeV (c), 35-40 GeV (d).
The lines show the results of fits containing a background component which is exponentially 
falling (dashed line) and a signal Gaussian component (dot-dashed line).  The total fit is
shown as a solid line.  The plots show the $\frac{S}{S+B}$ where the signal (S) is determined from the area
under the Gaussian within $\pm$2$\sigma$ of the mean and the total background (B)
includes both those jets reconstructed with a $>2\sigma$ energy mismatched and those 
which were not matched at all to a HIJING jet. Fit parameters are shown in Table~\ref{tab3}.}
\label{resp3}
\end{figure*}

\begin{figure*}
\centering
\subfigure[]{
\includegraphics[width=0.45\textwidth]{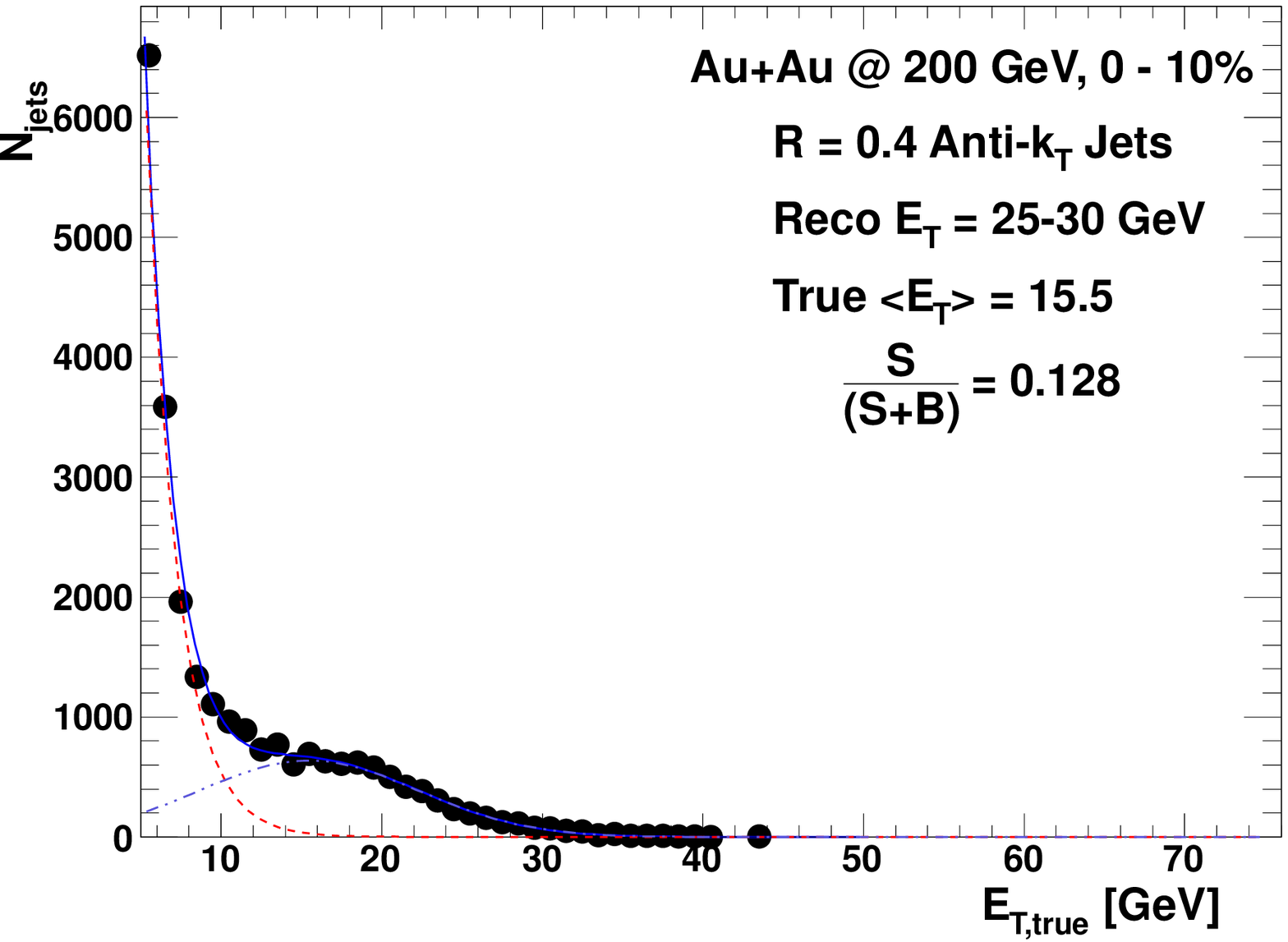}}
\hspace{0.05\textwidth}
\subfigure[]{
\includegraphics[width=0.45\textwidth]{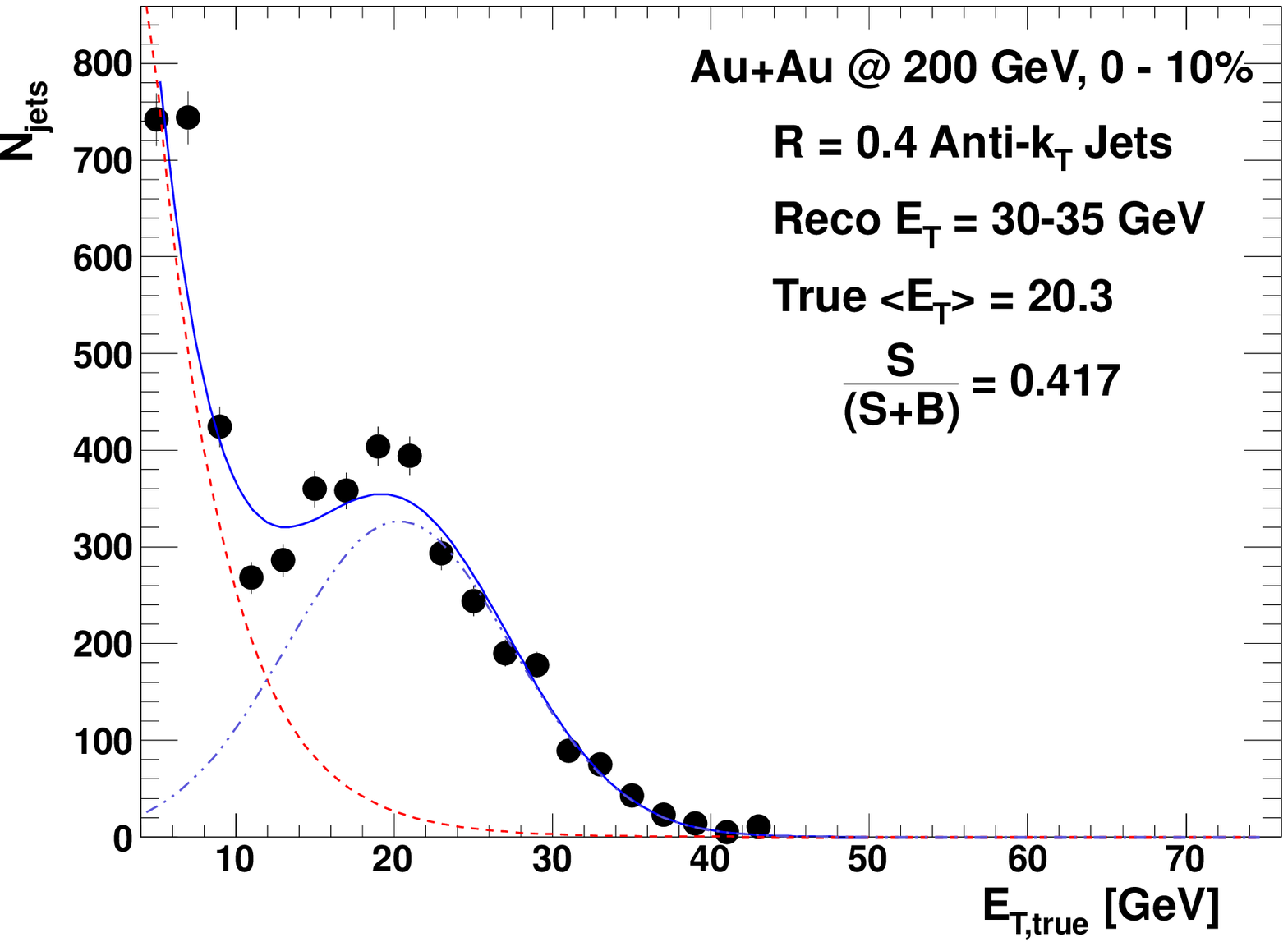}}
\subfigure[]{
\includegraphics[width=0.45\textwidth]{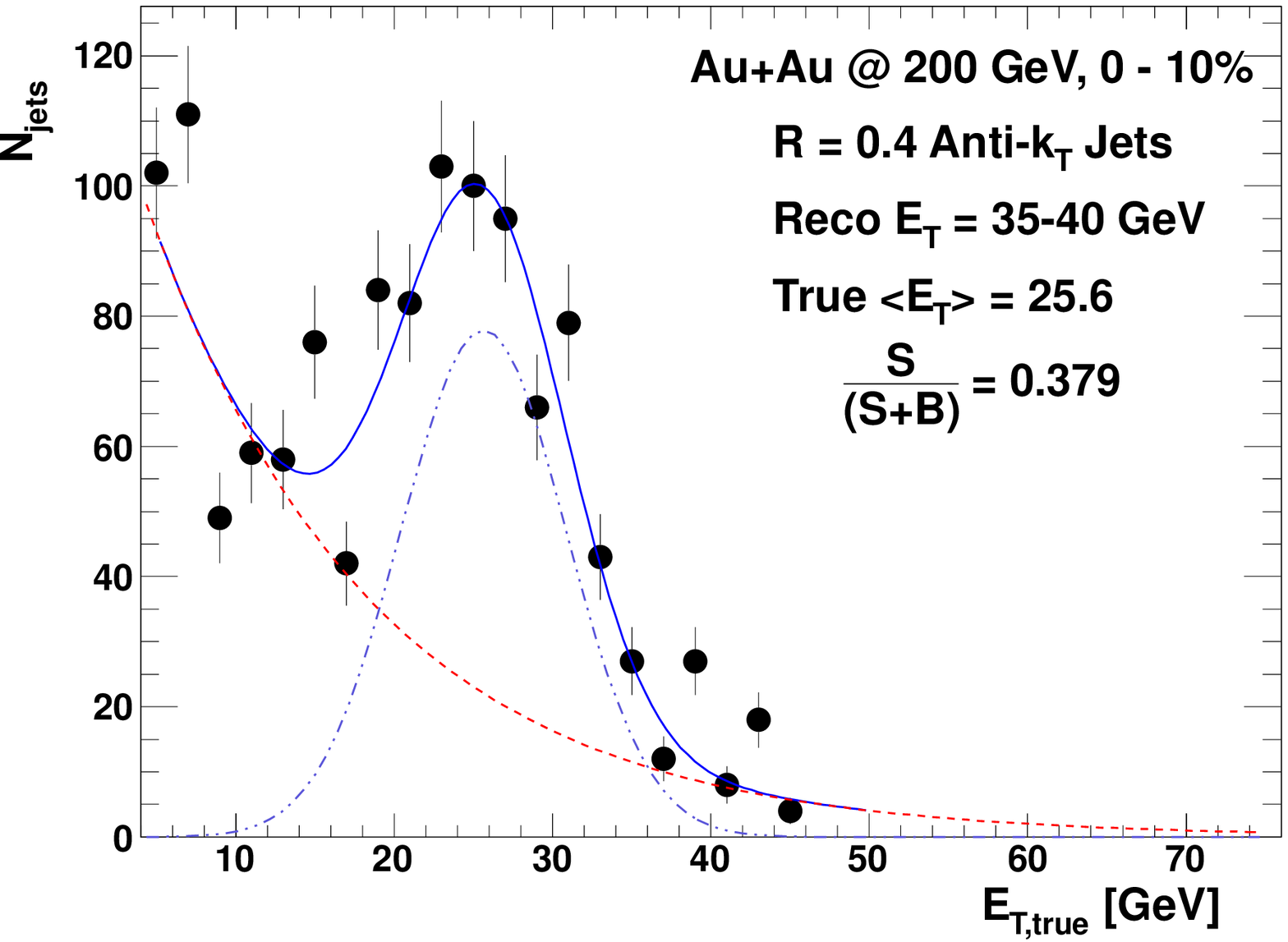}}
\hspace{0.05\textwidth}
\subfigure[]{
\includegraphics[width=0.45\textwidth]{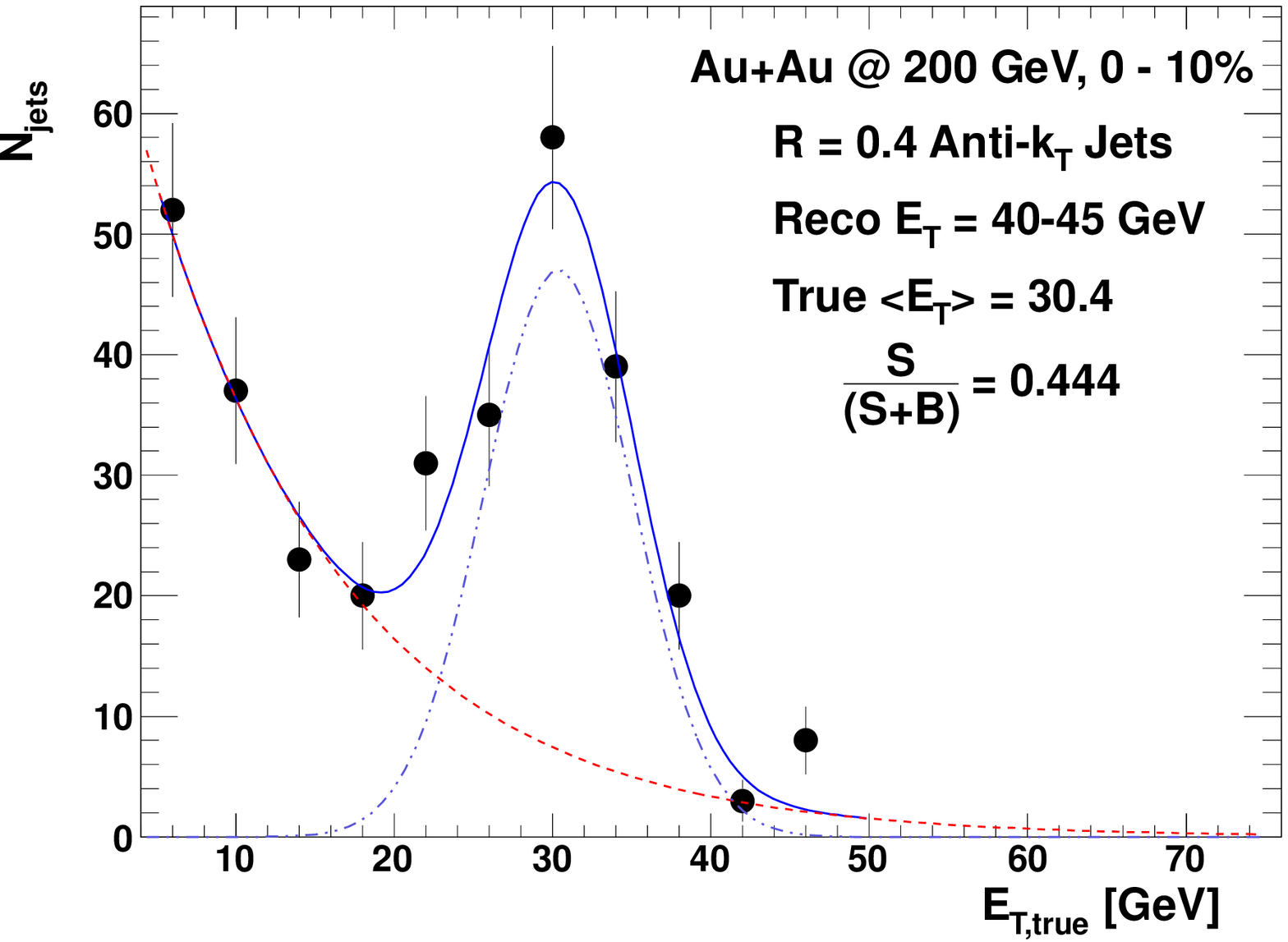}}
\caption{True \et for reconstructed jets anti-$k_{T}$ R = 0.4 for the reconstructed jet \et 
25-30 GeV (a), 30-35 GeV (b), 35-40 GeV (c), 40-45 GeV (d).
The lines show the results of fits containing a background component which is exponentially 
falling (dashed line) and a signal Gaussian component (dot-dashed line).  The total fit is
shown as a solid line.  The plots show the $\frac{S}{S+B}$ where the signal (S) is determined from the area
under the Gaussian within $\pm$2$\sigma$ of the mean and the total background (B)
includes both those jets reconstructed with a $>2\sigma$ energy mismatched and those 
which were not matched at all to a HIJING jet. Fit parameters are shown in Table~\ref{tab4}.}
\label{resp4}
\end{figure*}

\begin{table}[h]
\begin{tabular}{|c|c|c|c|} \hline
$<E_{T,reco}>$ (GeV) & $<E_{T,true}>$ (GeV) & $\sigma_{E_{T}}$ & $\frac{S}{S+B}$ \\ \hline
$17.2 \pm 0.00602$ & $14.6 \pm 0.0537$ & $3.77 \pm 0.0306$ & $0.211$\\ \hline 
$22.1 \pm 0.012$ & $18.6 \pm 0.051$ & $4.22 \pm 0.038$ & $0.745$\\ \hline 
$27.2 \pm 0.0271$ & $24.1 \pm 0.0815$ & $3.79 \pm 0.0526$ & $0.884$\\ \hline 
$32.3 \pm 0.0535$ & $29.0 \pm 0.152$ & $4.21 \pm 0.121$ & $0.925$\\ \hline 
\end{tabular} 
\caption{Jet parameters from fits to the plots in Figure~\ref{resp2} for R = 0.2 anti-$k_T$ jets with centrality 
from 0 - 10\%. $<E_{T,reco}>$ is the mean reconstructed jet \et within the 5  GeV wide bins.
$<E_{T,true}>$ and $\sigma_{E{T,true}}$ are the mean and width from the Gaussian component in
the fit and $\frac{S}{S+B}$ is the fraction of the area of the fit that is included in the 
Gaussian component rather than the exponential.}
\label{tab2} 
\end{table}

\begin{table}[h]
\begin{tabular}{|c|c|c|c|} \hline
$<E_{T,reco}>$ (GeV) & $<E_{T,true}>$ (GeV) & $\sigma_{E_{T}}$ & $\frac{S}{S+B}$ \\ \hline
$22.1 \pm 0.00748$ & $14.9 \pm 0.124$ & $5.59 \pm 0.0634$ & $0.174$\\ \hline 
$27.2 \pm 0.0169$ & $19.7 \pm 0.179$ & $5.83 \pm 0.144$ & $0.576$\\ \hline 
$32.2 \pm 0.0354$ & $23.7 \pm 0.261$ & $6.33 \pm 0.211$ & $0.802$\\ \hline 
$37.5 \pm 0.072$ & $29.6 \pm 0.329$ & $7.04 \pm 0.306$ & $0.908$\\ \hline 
\end{tabular} 
\caption{Jet parameters from fits to the plots in Figure~\ref{resp3} for R = 0.3 anti-$k_T$ jets with centrality 
from 0 - 10\%. $<E_{T,reco}>$ is the mean reconstructed jet \et within the 5  GeV wide bins.
$<E_{T,true}>$ and $\sigma_{E{T,true}}$ are the mean and width from the Gaussian component in
the fit and $\frac{S}{S+B}$ is the fraction of the area of the fit that is included in the 
Gaussian component rather than the exponential.}
\label{tab3} 
\end{table}

\begin{table}[h]
\begin{tabular}{|c|c|c|c|} \hline
$<E_{T,reco}>$ (GeV) & $<E_{T,true}>$ (GeV) & $\sigma_{E_{T}}$ & $\frac{S}{S+B}$ \\ \hline
$27.1 \pm 0.01$ & $15.5 \pm 0.188$ & $6.82 \pm 0.119$ & $0.128$\\ \hline 
$32.2 \pm 0.0222$ & $20.3 \pm 0.412$ & $7.06 \pm 0.258$ & $0.417$\\ \hline 
$37.3 \pm 0.0469$ & $25.6 \pm 0.357$ & $5.2 \pm 0.427$ & $0.379$\\ \hline 
$42.4 \pm 0.0987$ & $30.4 \pm 0.616$ & $4.69 \pm 0.726$ & $0.444$\\ \hline 
\end{tabular} 
\caption{Jet parameters from fits to the plots in Figure~\ref{resp4} for R = 0.4 anti-$k_T$ jets with centrality 
from 0 - 10\%. $<E_{T,reco}>$ is the mean reconstructed jet \et within the 5  GeV wide bins.
$<E_{T,true}>$ and $\sigma_{E{T,true}}$ are the mean and width from the Gaussian component in
the fit and $\frac{S}{S+B}$ is the fraction of the area of the fit that is included in the 
Gaussian component rather than the exponential.}
\label{tab4}
\end{table}

\begin{figure*}[h]
\centering
\includegraphics[width=\textwidth]{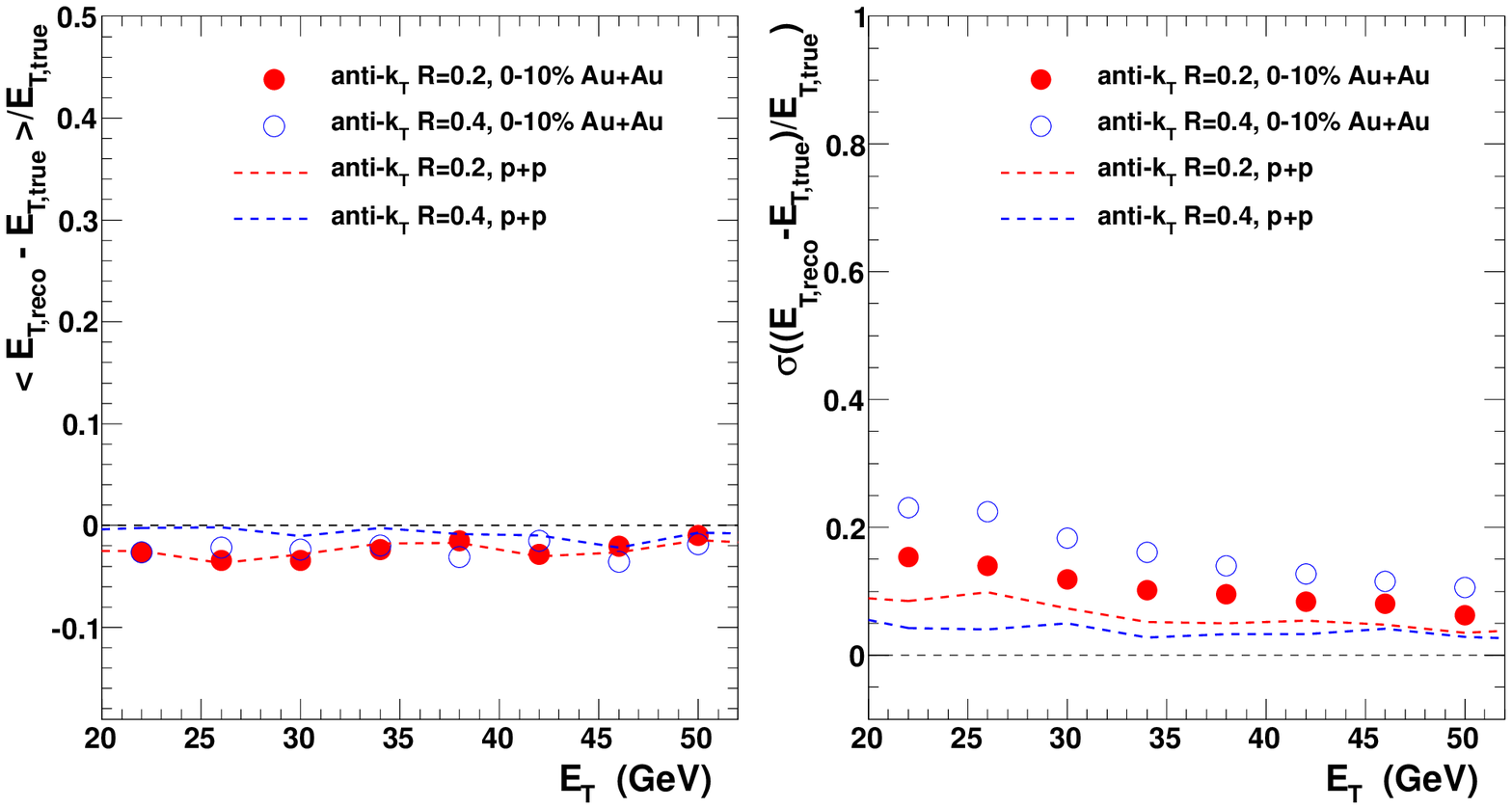}
\caption{Jet energy resolution as a function of the true jet energy for jets reconstructed
in 0-10\% central HIJING events.  Shown in the plot are anti-$k_T$ R = 0.2 (black), R = 0.3 (red)
and R = 0.4 (blue).}
\label{energy_res}
\end{figure*}

In addition to the fake jet contribution to the reconstructed jet sample it is also 
important to quantify the jet energy resolution and scale for our 
algorithm.  In order to do this we have embedded PYTHIA~\cite{pythia} (version 6.421) jets into
our HIJING events.  
One PYTHIA event with a high \pt dijet was embedded into every HIJING event.
The PYTHIA and reconstructed jets are required to obey the same
matching cut of $\Delta R<$ 0.25 as the fake jet study discussed above.  The jet energy
resolution and jet energy scale are shown in Figure~\ref{energy_res} 
 the anti-$k_T$ R parameters 0.2 and 0.4 for central HIJING events and PYTHIA events (not embedded into HIJING)
put into towers in the same manner as the HIJING events.  The jet energy
resolution improves with increasing jet energy and decreasing jet R as expected.
The energy scale, $\frac{\langle E_{T,reco} - E_{T,true} \rangle}{E_{T,true}}$ 
is within $\approx$5\% of zero for the anti-$k_{T}$ R parameters considered here.
The energy offset for the PYTHIA jets is due to the imposed tower segmentation.
For the purposes of this study we did not pursue further refinements.
A similar resolution evaluation has been done by the CMS collaboration~\cite{Chatrchyan:2011sx}.

\section{Conclusions}
We have performed a HIJING study of jet reconstruction using an iterative background
subtraction method and full calorimetric information.  We have shown that in this
case we are able to reconstruct the input HIJING jets with a large signal to background 
with $E_{T}>$ 20  GeV for R = 0.2 jets, 30  GeV for R = 0.3 jets and 40  GeV for R=0.4 jets.
The results presented here are obtained without any additionally rejection of fake jets, though
it is possible the reconstructed jet purities shown here could be further increased with fake
jet rejection of some kind.

This study was designed to evaluate the feasibility of purely calorimetric jet measurements at RHIC.
The results here are obtained using an ideal model of the detector and suggest promise for such measurements.
This study is of course limited in scope.  We have not taken into effect any detector effects
aside from geometrical acceptance and granularity.  
More detailed studies need to be done to demonstrate the suitability
of any particular detector design for jet measurements.

\section{Acknowledgments}
We would like to thank Yue Shi Lai for valuable discussions on the identification of jets in HIJING.
Funding has been provided from the Department of Energy under contract number DE-AC02-98CH10886
(A.M.S, A.F, D.P.M, C.H.P, and P.S) and grant numbers DE-FG02-96ER40988 (J.A.H, B.S.,M.vS)
and DE-FG02-03ER41244 (M.P.M, J.L.N, M.S).
\bibliography{fake_note}

\begin{thebibliography}{26}
\expandafter\ifx\csname natexlab\endcsname\relax\def\natexlab#1{#1}\fi
\expandafter\ifx\csname bibnamefont\endcsname\relax
  \def\bibnamefont#1{#1}\fi
\expandafter\ifx\csname bibfnamefont\endcsname\relax
  \def\bibfnamefont#1{#1}\fi
\expandafter\ifx\csname citenamefont\endcsname\relax
  \def\citenamefont#1{#1}\fi
\expandafter\ifx\csname url\endcsname\relax
  \def\url#1{\texttt{#1}}\fi
\expandafter\ifx\csname urlprefix\endcsname\relax\def\urlprefix{URL }\fi
\providecommand{\bibinfo}[2]{#2}
\providecommand{\eprint}[2][]{\url{#2}}

\bibitem[{\citenamefont{Gyulassy and Wang}(1994)}]{Gyulassy:1994ew}
\bibinfo{author}{\bibfnamefont{M.}~\bibnamefont{Gyulassy}} \bibnamefont{and}
  \bibinfo{author}{\bibfnamefont{X.}~\bibnamefont{Wang}},
  \bibinfo{journal}{Comput.~Phys.~Commun.} \textbf{\bibinfo{volume}{83}},
  \bibinfo{pages}{307} (\bibinfo{year}{1994}), \eprint{nucl-th/9502021}.

\bibitem[{\citenamefont{ATLAS}(2011)}]{atlas_note}
\bibinfo{author}{\bibnamefont{ATLAS}} (\bibinfo{collaboration}{ATLAS
  Collaboration}) (\bibinfo{year}{2011}),
  \urlprefix\url{cdsweb.cern.ch/record/1353220/files/ATLAS-CONF-2011-075.pdf}.

\bibitem[{\citenamefont{Aad et~al.}(2010)}]{Aad:2010bu}
\bibinfo{author}{\bibfnamefont{G.}~\bibnamefont{Aad}} \bibnamefont{et~al.}
  (\bibinfo{collaboration}{Atlas Collaboration}),
  \bibinfo{journal}{Phys.Rev.Lett.} \textbf{\bibinfo{volume}{105}},
  \bibinfo{pages}{252303} (\bibinfo{year}{2010}), \eprint{1011.6182}.

\bibitem[{\citenamefont{Chatrchyan et~al.}(2011)}]{Chatrchyan:2011sx}
\bibinfo{author}{\bibfnamefont{S.}~\bibnamefont{Chatrchyan}}
  \bibnamefont{et~al.} (\bibinfo{collaboration}{CMS Collaboration}),
  \bibinfo{journal}{Phys.Rev.} \textbf{\bibinfo{volume}{C84}},
  \bibinfo{pages}{024906} (\bibinfo{year}{2011}), \eprint{1102.1957}.

\bibitem[{\citenamefont{Adcox et~al.}(2002)}]{Adcox:2001jp}
\bibinfo{author}{\bibfnamefont{K.}~\bibnamefont{Adcox}} \bibnamefont{et~al.}
  (\bibinfo{collaboration}{PHENIX}), \bibinfo{journal}{Phys.~Rev.~Lett.}
  \textbf{\bibinfo{volume}{88}}, \bibinfo{pages}{022301}
  (\bibinfo{year}{2002}), \eprint{nucl-ex/0109003}.

\bibitem[{\citenamefont{Adams et~al.}(2003)}]{Adams:2003kv}
\bibinfo{author}{\bibfnamefont{J.}~\bibnamefont{Adams}} \bibnamefont{et~al.}
  (\bibinfo{collaboration}{STAR Collaboration}),
  \bibinfo{journal}{Phys.Rev.Lett.} \textbf{\bibinfo{volume}{91}},
  \bibinfo{pages}{172302} (\bibinfo{year}{2003}), \eprint{nucl-ex/0305015}.

\bibitem[{\citenamefont{Adare et~al.}(2008{\natexlab{a}})}]{Adare:2008qa}
\bibinfo{author}{\bibfnamefont{A.}~\bibnamefont{Adare}} \bibnamefont{et~al.}
  (\bibinfo{collaboration}{PHENIX}), \bibinfo{journal}{Phys. Rev. Lett.}
  \textbf{\bibinfo{volume}{101}}, \bibinfo{pages}{232301}
  (\bibinfo{year}{2008}{\natexlab{a}}), \eprint{0801.4020}.

\bibitem[{\citenamefont{Adams et~al.}(2006)}]{Adams:2006yt}
\bibinfo{author}{\bibfnamefont{J.}~\bibnamefont{Adams}} \bibnamefont{et~al.}
  (\bibinfo{collaboration}{STAR Collaboration}),
  \bibinfo{journal}{Phys.Rev.Lett.} \textbf{\bibinfo{volume}{97}},
  \bibinfo{pages}{162301} (\bibinfo{year}{2006}), \eprint{nucl-ex/0604018}.

\bibitem[{\citenamefont{Adare et~al.}(2008{\natexlab{b}})}]{Adare:2008cqb}
\bibinfo{author}{\bibfnamefont{A.}~\bibnamefont{Adare}} \bibnamefont{et~al.}
  (\bibinfo{collaboration}{PHENIX Collaboration}), \bibinfo{journal}{Phys.Rev.}
  \textbf{\bibinfo{volume}{C78}}, \bibinfo{pages}{014901}
  (\bibinfo{year}{2008}{\natexlab{b}}), \eprint{0801.4545}.

\bibitem[{\citenamefont{Abelev et~al.}(2009)}]{Abelev:2009qa}
\bibinfo{author}{\bibfnamefont{B.}~\bibnamefont{Abelev}} \bibnamefont{et~al.}
  (\bibinfo{collaboration}{STAR Collaboration}), \bibinfo{journal}{Phys.Rev.}
  \textbf{\bibinfo{volume}{C80}}, \bibinfo{pages}{064912}
  (\bibinfo{year}{2009}), \eprint{0909.0191}.

\bibitem[{\citenamefont{Adare et~al.}(2010)}]{Adare:2010ry}
\bibinfo{author}{\bibfnamefont{A.}~\bibnamefont{Adare}} \bibnamefont{et~al.}
  (\bibinfo{collaboration}{PHENIX}), \bibinfo{journal}{Phys.~Rev.~Lett.}
  \textbf{\bibinfo{volume}{104}}, \bibinfo{pages}{252301}
  (\bibinfo{year}{2010}), \eprint{1002.1077}.

\bibitem[{\citenamefont{Wang et~al.}(1996)\citenamefont{Wang, Huang, and
  Sarcevic}}]{Wang:1996yh}
\bibinfo{author}{\bibfnamefont{X.-N.} \bibnamefont{Wang}},
  \bibinfo{author}{\bibfnamefont{Z.}~\bibnamefont{Huang}}, \bibnamefont{and}
  \bibinfo{author}{\bibfnamefont{I.}~\bibnamefont{Sarcevic}},
  \bibinfo{journal}{Phys.Rev.Lett.} \textbf{\bibinfo{volume}{77}},
  \bibinfo{pages}{231} (\bibinfo{year}{1996}), \eprint{hep-ph/9605213}.

\bibitem[{\citenamefont{Steinberg and Collaboration}(2011)}]{Steinberg:2011qq}
\bibinfo{author}{\bibfnamefont{P.}~\bibnamefont{Steinberg}} \bibnamefont{and}
  \bibinfo{author}{\bibfnamefont{A.}~\bibnamefont{Collaboration}}
  (\bibinfo{year}{2011}), \eprint{1110.3352}.

\bibitem[{\citenamefont{Majumder and Van~Leeuwen}(2010)}]{Majumder:2010qh}
\bibinfo{author}{\bibfnamefont{A.}~\bibnamefont{Majumder}} \bibnamefont{and}
  \bibinfo{author}{\bibfnamefont{M.}~\bibnamefont{Van~Leeuwen}}
  (\bibinfo{year}{2010}), \eprint{1002.2206}.

\bibitem[{\citenamefont{Aamodt et~al.}(2011)}]{Aamodt:2010cz}
\bibinfo{author}{\bibfnamefont{K.}~\bibnamefont{Aamodt}} \bibnamefont{et~al.}
  (\bibinfo{collaboration}{ALICE Collaboration}),
  \bibinfo{journal}{Phys.Rev.Lett.} \textbf{\bibinfo{volume}{106}},
  \bibinfo{pages}{032301} (\bibinfo{year}{2011}), \eprint{1012.1657}.

\bibitem[{\citenamefont{Lai}(2009)}]{Lai:2009ai}
\bibinfo{author}{\bibfnamefont{Y.}~\bibnamefont{Lai}}
  (\bibinfo{collaboration}{PHENIX}) (\bibinfo{year}{2009}), \eprint{0911.3399}.

\bibitem[{\citenamefont{Jacobs}(2010)}]{Jacobs:2010wq}
\bibinfo{author}{\bibfnamefont{P.}~\bibnamefont{Jacobs}}
  (\bibinfo{collaboration}{STAR Collaboration}) (\bibinfo{year}{2010}),
  \eprint{1012.2406}.

\bibitem[{\citenamefont{Cacciari et~al.}(2011)\citenamefont{Cacciari, Salam,
  and Soyez}}]{Cacciari:2011tm}
\bibinfo{author}{\bibfnamefont{M.}~\bibnamefont{Cacciari}},
  \bibinfo{author}{\bibfnamefont{G.~P.} \bibnamefont{Salam}}, \bibnamefont{and}
  \bibinfo{author}{\bibfnamefont{G.}~\bibnamefont{Soyez}},
  \bibinfo{journal}{Eur.Phys.J.} \textbf{\bibinfo{volume}{C71}},
  \bibinfo{pages}{1692} (\bibinfo{year}{2011}), \eprint{1101.2878}.

\bibitem[{\citenamefont{Abelev et~al.}(2012)}]{Abelev:2012ej}
\bibinfo{author}{\bibfnamefont{B.}~\bibnamefont{Abelev}} \bibnamefont{et~al.}
  (\bibinfo{collaboration}{ALICE Collaboration}) (\bibinfo{year}{2012}),
  \eprint{1201.2423}.

\bibitem[{\citenamefont{Vogelsang}()}]{Vogelsang:NLO}
\bibinfo{author}{\bibfnamefont{W.}~\bibnamefont{Vogelsang}},
  \bibinfo{note}{private communication}.

\bibitem[{\citenamefont{PHENIX}(2010)}]{decadal_plan}
\bibinfo{author}{\bibnamefont{PHENIX}} (\bibinfo{year}{2010}),
  \urlprefix\url{{http://www.phenix.bnl.gov/phenix/WWW/docs/decadal/2010/phenix_decadal10_full_refs.pdf}}.

\bibitem[{\citenamefont{Cacciari et~al.}(2008)\citenamefont{Cacciari, Salam,
  and Soyez}}]{Cacciari:2008gp}
\bibinfo{author}{\bibfnamefont{M.}~\bibnamefont{Cacciari}},
  \bibinfo{author}{\bibfnamefont{G.}~\bibnamefont{Salam}}, \bibnamefont{and}
  \bibinfo{author}{\bibfnamefont{G.}~\bibnamefont{Soyez}},
  \bibinfo{journal}{JHEP} \textbf{\bibinfo{volume}{0804}}, \bibinfo{pages}{063}
  (\bibinfo{year}{2008}), \eprint{0802.1189}.

\bibitem[{\citenamefont{Cacciari and Salam}(2006)}]{Cacciari:2005hq}
\bibinfo{author}{\bibfnamefont{M.}~\bibnamefont{Cacciari}} \bibnamefont{and}
  \bibinfo{author}{\bibfnamefont{G.~P.} \bibnamefont{Salam}},
  \bibinfo{journal}{Phys.Lett.} \textbf{\bibinfo{volume}{B641}},
  \bibinfo{pages}{57} (\bibinfo{year}{2006}), \eprint{hep-ph/0512210}.

\bibitem[{\citenamefont{Masera et~al.}(2009)\citenamefont{Masera, Ortona,
  Poghosyan, and Prino}}]{Masera:2009zz}
\bibinfo{author}{\bibfnamefont{M.}~\bibnamefont{Masera}},
  \bibinfo{author}{\bibfnamefont{G.}~\bibnamefont{Ortona}},
  \bibinfo{author}{\bibfnamefont{M.}~\bibnamefont{Poghosyan}},
  \bibnamefont{and} \bibinfo{author}{\bibfnamefont{F.}~\bibnamefont{Prino}},
  \bibinfo{journal}{Phys.~Rev.} \textbf{\bibinfo{volume}{C79}},
  \bibinfo{pages}{064909} (\bibinfo{year}{2009}).

\bibitem[{\citenamefont{Adare et~al.}(2011)}]{Adare:2011tg}
\bibinfo{author}{\bibfnamefont{A.}~\bibnamefont{Adare}} \bibnamefont{et~al.}
  (\bibinfo{collaboration}{PHENIX Collaboration}),
  \bibinfo{journal}{Phys.Rev.Lett.} \textbf{\bibinfo{volume}{107}},
  \bibinfo{pages}{252301} (\bibinfo{year}{2011}), \eprint{1105.3928}.

\bibitem[{\citenamefont{Sjostrand et~al.}(2001)}]{pythia}
\bibinfo{author}{\bibfnamefont{T.}~\bibnamefont{Sjostrand}}
  \bibnamefont{et~al.}, \bibinfo{journal}{Comput.~Phys.~Commun.}
  \textbf{\bibinfo{volume}{135}}, \bibinfo{pages}{238} (\bibinfo{year}{2001}),
  \eprint{hep-ph/0010017}.

\end{thebibliography}
\end{document}